\documentclass[11pt,a4paper]{article}
\usepackage{graphicx}
\usepackage{jcappub}
\usepackage{float}
\usepackage{lscape}
\usepackage{pdflscape}
\usepackage{indentfirst}
\newcommand{\be}{\begin{equation}}
\newcommand{\ee}{\end{equation}}

\newcommand{\ba}{\begin{eqnarray}}
\newcommand{\ea}{\end{eqnarray}}

\usepackage{float}

\newcommand{\bege}{\begin{equation}}
\newcommand{\bpartial}{\mathop{\partial\kern -4pt\raisebox{.8pt}{$|$}}}
\newcommand{\enge}{\end{equation}}
\newcommand{\beq}{\begin{eqnarray}}
\newcommand{\benu}{\begin{enumerate}}

\newcommand{\enu}{\end{enumerate}}
\newcommand{\eeq}{\end{eqnarray}}

\begin{document}

\title{{Cosmological  reconstruction and {\it Om} diagnostic  analysis of Einstein-Aether Theory}}

\author[a]{Antonio Pasqua}
\affiliation[a]{Department of Physics,
University of Trieste, Via Valerio, 2 34127 Trieste, Italy.}
 \author[b]{Surajit Chattopadhyay}
\affiliation[b]{Pailan College of Management and Technology, Bengal Pailan Park, Kolkata-700 104, India.}
 \author[c]{Davood Momeni}
 \affiliation[c]{Eurasian International Center for Theoretical Physics and Department of
General \& Theoretical Physics, Eurasian National University,
Astana 010008, Kazakhstan}
 \author[d]{Muhammad Raza}
 \affiliation[d]{Department of Mathematics, COMSATS Institute of Information Technology, Sahiwal 57000, Pakistan and State Key Lab of Modern Optical Instrumentation,
Centre for Optical and Electromagnetic Research,
Department of Optical Engineering, Zhejiang University, Hangzhou 310058, China}
 \author[e]{Ratbay Myrzakulov}
 \affiliation[e]{Eurasian International Center for Theoretical Physics and Department of
General \& Theoretical Physics, Eurasian National University,
Astana 010008, Kazakhstan}
\author[f,g]{Mir Faizal}
 \affiliation[f]{Irving K. Barber School of Arts and Sciences, \\University of British Columbia - Okanagan,\\  Kelowna, British Columbia V1V 1V7, Canada}
 \affiliation[g]{ Department of Physics and Astronomy, University of Lethbridge, Lethbridge, Alberta T1K 3M4, Canada}
\emailAdd{toto.pasqua@gmail.com}
\emailAdd{surajcha@iucaa.ernet.in}
\emailAdd{d.momeni@yahoo.com}
\emailAdd{mraza@zju.edu.cn}
\emailAdd{e-mail:rmyrzakulov@gmail.com}
\emailAdd{e-mail:mirfaizalmir@gmail.com}

\begin{abstract}
{In this paper, we will analyse the cosmological models in Einstein-aether gravity, which is a modified theory of gravity in which
a time-like vector field breaks the Lorentz symmetry. We will use this formalism to
analyse different cosmological models with different behavior of the scale factor. In this analysis, we will use a  certain functional dependence of the
dark energy on the Hubble parameter. It will be demonstrated that the aether vector field has a non-trivial effect on these cosmological models.
We will also  perform the    \emph{Om} diagnostic   in Einstein-aether gravity. Thus, we will fit parameters  of the cosmological models  using recent observational data.}
\end{abstract}
\flushbottom

\maketitle

\section{Introduction}

The Lorentz symmetry is one of the most important symmetries in nature, and all particle physics experiments have
demonstrated that this symmetry is not broken at the scale at which such experiments are performed.
However, it is predicted from quantum gravity that the Lorentz symmetry should break down at Planck scale, where
even the manifold structure of spacetime breaks down due to quantum fluctuations. In fact, almost all  approaches
to quantum gravity predict that the local Lorentz symmetry of spacetime only exists in some  IR limit of the theory.
So, the Lorentz symmetry is expected to break in the UV limit. It may be noted that it has been explicitly demonstrated
that such a breaking of Lorentz symmetry in the UV limit occur in  the discrete spacetime \cite{Hooft},
models based on string field theory \cite{Samuel1}, spacetime foam \cite{Ellis}, spin-network in loop quantum gravity (LQG)
\cite{Gambini}, non-commutative geometry
\cite{Carroll,FaizalMPLA}, and ghost condensation in perturbative quantum gravity
\cite{FaizalJPA}. As the Lorentz symmetry fixes the form of the energy-momentum dispersion relation, the breaking of Lorentz symmetry
in the UV limit, will also lead to a modification of the energy-momentum dispersion relation in the UV limit.
In fact, there are indications from the
Greisen-Zatsepin-Kuzmin limit (GZK limit)  t the usual
energy-momentum relation will get modified in the UV limit
\cite{Greisen,Zatsepin}.  The Pierre Auger Collaboration and the High Resolution
Fly's Eye (HiRes) experiment have confirmed earlier results of
the GZK cutoff \cite{Abraham}. So, it is possible that the Lorentz symmetry will break in the UV limit, and only occur as an effective
symmetry in the IR limit. Thus, it is important to construct a  theory, such that it will reproduce the general relativity in the IR limit,
and break the local Lorentz symmetry in the UV limit. Such a theory has been constructed by using different Lifshitz scaling for space and time,
and this theory is called the   Horava-Lifshitz gravity \cite{HoravaPRD,HoravaPRL}. This original proposal for the Horava-Lifshitz gravity improves the
renormalization of   gravity, as it differs from general relativity in the UV limit. However, there are several problems associated with this proposal, and
these include the problems associated with
instabilities,
overconstrained evolution, and  strong coupling at low energies \cite{1}-\cite{1d}. These problems occur due to a  badly behaved scalar mode of gravity,
which is produced by  the presence of a nondynamical spatial
foliation in the action. To resolve this problem, an extension of  Horava-Lifshitz gravity called the BPSH theory has been proposed \cite{2}.
It has been demonstrated that the BPSH theory is equivalent to general relativity coupled to a dynamical unit timelike vector field \cite{2a}.
Here the vector is restricted in the action to be hypersurface orthogonal.

The  phenomenology and observational constraints on the coupling parameters of Einstein-aether gravity have been studied \cite{4a}.
 It may also be noted that constraints on Einstein-aether gravity from binary  pulsars have also been discussed \cite{bi}.
In this work, the consequences of  Lorentz symmetry, which occur in Einstein-aether gravity,  on the
orbital evolution of binary pulsars. In the focus of this study was on the dissipative effects in such a process.
It was observed that the breaking of the Lorentz symmetry  modified such  effects.
Thus, the   orbital dynamics of binary pulsars was also modified in Einstein-aether gravity. Such a modification
  causes the emission of dipolar radiation, and this made    the orbital separation decrease faster than in general relativity.
The quadruple component of the emission was also modified.
The orbital evolution depends critically on the sensitivities of the stars,
as this  measure how their binding energies depend on the motion relative to the preferred frame.
In this study such  sensitivities have also been  numerically  calculated in Einstein-aether gravity. These
 predictions have been compared with  observations,  and this has been used to set    constraints on Einstein-aether gravity.

It has been demonstrated that
 the Einstein-aether theory  can be analysed   in the framework of the metric-affine   gravity  \cite{5a}.
 Such a formalism resembles the  gauge theory of theory. In this formalism,  the aether vector field is related
 to certain post-Riemannian nonmetricity pieces contained in an independent linear connection of spacetime.
 Black hole solution have also been studied in the Einstein-aether gravity. It has been demonstrated that
  the deviations   from the Schwarzschild metric are typically only   a few percent for most of the explored parameter regions,
  and this   makes it  difficult to observe with electromagnetic probes, but they can be detected using
 gravitational wave detectors \cite{6a}. As gravitational wave detectors are going to be used extensively in future astronomy, it is interesting to study
 the implications of   Einstein-aether gravity.

As the Einstein-aether gravity, introduce a time like vector field, they are expected to modify the cosmological evolution of the universe.
In fact, various different solution for the
accelerating universe  in  the  Einstein-aether gravity have been studied \cite{as}.
These solutions have been used to  analyse the inflationary behaviour of  the early universe and late-time cosmological acceleration.
It has been demonstrated that
the aether field produces  accelerated expansion in situations where inflation
would not occur in general relativity. Hence, the aether field can effect the inflation in a very non-trivial way.
 The cosmological evolution of cosmological models based on Einstein-aether gravity  with power-law potential have also been studied \cite{po}.
 The cosmological models have also been studied in the Einstein-aether gravity coupled to a Galileon type scalar field \cite{ga}.
 It was observed that in such models, the universe experiences a late time acceleration, for   pressure-less baryonic matter.
 It may be noted that gravitational wave can be used to analyse the cosmological aspects of Einstein-aether gravity. This is because
   it has been demonstrated that for cosmological models based   Einstein-aether gravity,  a direct correspondence exists between perfect fluids
   carrying anisotropic stress and a modification in the propagation of gravitational waves \cite{gw}.  As
   the anisotropic stress can be measured in a model-independent manner,
   the  gravitational waves can be used to obtain    constraints on the cosmological models in the  Einstein-aether gravity.
 Even though several studies have been done on the Einstein-aether gravity, it is important to perform an extensive study of how
 this modification of general relativity can change cosmological models, and how it fits with the current data. This is important
 as many aspects of Einstein-aether gravity can be detected using gravitational waves, and in near future, it is expected that  the gravitational wave   will be
 used  to study many of these interesting  phenomena. So, in this paper, we perform a detail study on the modification of different cosmological models from
 Einstein-aether gravity. These models have been studied in general relativity, and we will,  and we will analyse  them in the Einstein-aether gravity.
So, in this paper,  we will  first  use the reconstruction technique to find
some viable forms for Einstein-aether gravity. We obtain the expressions of the modified Friedmann equations
and  from these equations, we can find the effective density and the effective pressure for the Einstein-aether gravity. We
will also fit the model with observational data. This will be done by using  the cosmographic analysis involving  the {\it Om} parametrization.
We will also use the SnIa, BAO and Hubble
data to find the $1\sigma$ and $2\sigma$ contours for density parameter $\Omega_m$   arising from the Sne Ia + BAO.

\section{Einstein-aether  Gravity}
In this section, we will review  the main features of the Einstein-aether    gravity.
It may be noted that Einstein-aether gravity is equivalent to the BPSH generalization to the Horava-Lifshitz gravity \cite{2a},
and so it breaks the  Lorentz symmetry. However,    the cosmology described by this theory  would still be described by the standard Friedmann equations with an additional
matter contribution \cite{as}. This is because the breaking of Lorentz symmetry occurs due to a time-like vector field in the Einstein-aether theory,
and so   the cosmological effects  can be obtained by analysing the correction to the standard Friedmann equations from this additional time-like vector field.
The action $S$ of the Einstein-aether    gravity    is given by \cite{48,49},
\begin{eqnarray}
S = \int d^4x \sqrt{-g}\left[\frac{R}{4\pi G} + L_{EA} + L_m    \right], \label{1}
\end{eqnarray}
where $L_{EA}$ indicates the Lagrangian density of the aether vector field,
while $L_m$ indicates the Lagrangian density of the usual matter fields.
The Lagrangian density of the aether vector field $L_{EA}$  can be expressed as
\begin{eqnarray}
L_{EA} &=& \frac{M^2}{16\pi G}F \left( K \right) + \frac{1}{16\pi G}\lambda \left( A^aA_a+1  \right),  \label{2}     \\
K &=& M^{-2} K^{ab}_{cd}\nabla_a A^c \nabla_b A^d,   \label{3}        \\
K^{ab}_{cd}&=&c_1 g^{ab}g_{cd}+c_2\delta_c^a \delta_d^b + c_3\delta_d^a \delta_c^b, \ \ a,b=0,1,2,3\label{4}
\end{eqnarray}
where $c_1$, $c_2$ and $c_3$ are three dimensionless constant parameters, $M$ is a coupling constant parameter,
$\lambda$ is a Lagrangian multiplier, and $A^a$ is a contravariant vector. Here    $F\left( K \right)$ is an arbitrary function of the parameter $K$,
and  a function of  the Hubble parameter $H$.\\
Now using  Eq. (\ref{1}),   the   field equations, for this theory, can be written as \cite{48,49}
\begin{eqnarray}
G_{ab} &=& T_{ab}^{Einstein-aether   } + 8\pi G T_{ab}^m,   \label{5}    \\
\nabla_a \left(F' J^a_{\,b}   \right) &=& 2\lambda A_b,\label{6}
\end{eqnarray}
where
\begin{eqnarray}
F'&=&\frac{dF}{dK}, \label{7}  \\
J^a_{\,b}= &-& 2K^{ad}_{\, \, bc} \nabla _d A^c. \label{8}
\end{eqnarray}
Here $T_{ab}^m$ is the energy-momentum tensors for the matter field, and $T_{ab}^{Einstein-aether   }$ is the energy-momentum tensor
for the aether vector field,
\begin{eqnarray}
T_{ab}^m &=& \left( p+\rho \right)u_a u_b +pg_{ab},\label{9} \\
T_{ab}^{Einstein-aether   } &=& \frac{1}{2}\nabla_d \left[\left( J_a^{\,d}A_b- J^d_{\,a}A_b
   - J_{\left(ab\right)}A^d   \right)F'\right]  \nonumber \\ && - Y_{\left(ab\right)}F' + \frac{1}{2}g_{ab}M^2F + \lambda A_a A_b.\label{10}
\end{eqnarray}
Here $\rho$ is   the energy density and   $p$ is the pressure of the matter. Also
$u_a$ is defined as $u_a = \left( 1,0,0,0 \right)$, and  so it represents the four-velocity vector of the fluid, $A^a =  \left( 1,0,0,0 \right)$. It may be noted that it
is represented by  a time-like unitary vector. Now we define $Y_{ab}$  as
\begin{eqnarray}
Y_{ab} = -c_1 \left[ \left( \nabla_d A_a \right)\left( \nabla^d A_b \right) - \left( \nabla_a A_d \right)\left( \nabla_a A^d \right)\right].   \label{11}
\end{eqnarray}
The subscript $\left( ab\right)$ indicates a symmetry with respect to the two indices.

We consider a Friedmann-Robertson-Walker (FRW) metric,  which is given by
\begin{eqnarray}
ds^2 = -dt^2 +a^2\left(t\right)\left[\frac{dr^2}{1-kr^2} + r^2\left(d\theta^2 + \sin^2 \theta d\phi^2   \right)   \right], \label{12}
\end{eqnarray}
where $a\left(t\right)$ represents the scale factor (which gives useful information about the expansion rate of the universe),
$t$ is the cosmic time, and $k$ is the curvature parameter. Here  its  value can be $-1$, $0$ or $+1$ corresponding
to an open, a flat, or a closed Universe.   The range of
  $\theta$ and $\phi$ are   $0\leq \theta \leq \pi$ and $0\leq \phi \leq 2\pi$.
The four coordinates $\left(r , t, \theta, \phi  \right)$ are also known as co-moving coordinates.

Using Eqs. (\ref{3}) and (\ref{4}), we can easily obtain the following general expression for the parameter $K$ \cite{48,49},
\begin{eqnarray}
K = \frac{3\varepsilon H^2}{M^2}, \label{13}
\end{eqnarray}
where $\varepsilon$ is a constant parameter while $H=\dot{a}/a$ is the Hubble parameter.
So, using  Eq. (\ref{5}), we obtain the   Friedmann equations modified by the  Einstein-aether    gravity,
\begin{eqnarray}
\varepsilon \left(-F' + \frac{F}{2K}   \right)H^2 + \left( H^2 + \frac{k}{a^2}  \right) &=& \frac{8\pi G}{3}\rho,  \label{14} \\
\varepsilon \frac{d}{dt}\left( HF'  \right) + \left( -2\dot{H} + \frac{2k}{a^2}  \right) &=& 8\pi G \left( p+ \rho  \right).   \label{15}
\end{eqnarray}
 The conservation equation can now be written as
\begin{eqnarray}
\dot{\rho} + 3H \left( p+ \rho  \right) =0, \label{16}
\end{eqnarray}
where the $\dot{\rho}$ indicates a temporal  derivative of $\rho$.

We   denote  the effective energy density in Einstein-aether    gravity by $\rho_{EA}$,  and the effective pressure in Einstein-aether    gravity
by $p_{EA}$. So, we can rewrite Eqs. (\ref{14}) and (\ref{15}) as
\begin{eqnarray}
 \left( H^2 + \frac{k}{a^2}  \right) &=& \frac{8\pi G}{3}\rho + \frac{1}{3}\rho_{EA}, \label{16} \\
\left( -2\dot{H} + \frac{2k}{a^2}  \right) &=& 8\pi G \left( p+ \rho  \right) + \left( \rho_{EA} + p_{EA}  \right).\label{17}
\end{eqnarray}
Therefore, comparing Eqs. (\ref{14}) and (\ref{15}) with Eqs. (\ref{16}) and (\ref{17}),
we can write
\begin{eqnarray}
\rho_{EA} &=& 3\varepsilon H^2 \left(F' - \frac{F}{2K}   \right), \label{18} \\
p_{EA} &=& -3\varepsilon H^2 \left(F' - \frac{F}{2K}   \right) - \varepsilon \left( \dot{H}F' +H\dot{F}'   \right)
\nonumber \\ &=& - \rho_{EA}  - \frac{\dot{\rho}_{EA}}{3H}.\label{19}
\end{eqnarray}
Using  Eq. (\ref{18}), we obtain
\begin{eqnarray}
F' - \frac{F}{2K}   = \frac{\rho_{EA} }{3\varepsilon H^2}. \label{18a}
\end{eqnarray}
This  is equivalent to the following  master  equation (using the expression of $3\varepsilon H^2$ derived from Eq. (\ref{13})),
\begin{eqnarray}
F' - \frac{F}{2K}   = \frac{\rho_{EA} }{KM^2}. \label{18b}
\end{eqnarray}
Here that a prime $'$ indicates a derivative with respect to $K$, and so we have   $F' = \frac{dF}{dK}$.\\
Using the expressions for $\rho_{EA}$ and $p_{EA}$ given by  Eqs. (\ref{18}) and (\ref{19}),
we obtain that the EoS parameter   $\omega_{EA}$  for  the Einstein-aether  gravity,
\begin{eqnarray}
\omega_{EA} = \frac{p_{EA}}{\rho_{EA}} = -1 - \frac{\left( \dot{H}F' +H\dot{F}'   \right)}{3H^2\left(F' - \frac{F}{2K}   \right)}. \label{20}
\end{eqnarray}

\section{Models for Dark Energy }
The   astrophysical  data  obtained from   distant  Ia  supernovae,  large  scale  structure,  baryon acoustic oscillations, weak lensing  and
cosmic microwave background indicate the existence of dark energy \cite{data,data2,data3,data4,data5,data6,data7,data8,data9,data10}.
In this paper, we will analyse the dark energy models using Einstein-aether gravity.
The
effective density and the effective pressure produced  by Einstein-aether    gravity can be used  to generate dark energy,
if the condition $\rho_{EA} + 3 p_{EA} <0$ is satisfied (i.e., if the
strong energy condition is violated). Thus, we obtain
\begin{eqnarray}
 2H^2 \left(F' - \frac{F}{2K}   \right) > - \left( \dot{H}F' +H\dot{F}'   \right).
\end{eqnarray}
This is the equation can be used to analyse dark energy  in Einstein-aether    gravity.
So, to analyse the effect of dark energy on  cosmological models in the Einstein-aether gravity, we can use
 $F\left(K\right)$ and $\omega_{Einstein-aether  }$. It may be noted that the  modified
  effective Friedmann  equation can be written as
\begin{eqnarray}
8\pi G\rho_{dark energy }=\Sigma_{\Sigma n_{i}=n} A(X,Y,...)\frac{\partial^n f(X,Y,...)}{\partial X^{n_1} \partial Y^{n_2}...} \label{partial}.
\end{eqnarray}
where  $f(X,Y,...)$ is the matter part $K$ action. For a model of dark energy ,
for example,  the holographic dark energy, we have  $\rho_{dark energy }=\rho_{dark energy }(H,\dot{H},..)$. We can also write it as
$\rho_{dark energy }=\rho_{dark energy }(X,Y,...)$. So, if we can solve the  partial differential equations for $f(X,Y,...)$,  we can
obtain the effect of dark energy on such cosmological models.

The generalized Nojiri-Odintsov Holographic dark energy  models can be used to analyse the dependence of the dark energy on
the Hubble parameter \cite{hep-th/0506212}.  In fact, using the  Granda-Oliveros model \cite{lgo}, which is a specific kind of
Nojiri-Odintsov Holographic dark energy  model, we can write,
\begin{eqnarray}
L_{GO}= \left(\alpha H^2 + \beta \dot{H}   \right)^{-\frac{1}{2}}, \label{deflgo}
\end{eqnarray}
where $H={\dot{a}}/{a}$ is the Hubble parameter, and $\dot{H}$ is the temporal derivative of $H$.
This model is characterized by two constant parameters $\alpha$ and $\beta$.
 As the  dark energy  dominates the present cosmological epoch, and its contribution to cosmological epoch near
 the Big Bang was negligible (i.e. the amount of dark energy  increased with the expansion of the universe),   the energy density can be assumed to be a
  function of the Hubble parameter $H$ and its temporal  derivative. Such a dependence is characterized by these two parameters.
  There are other physical reasons which motivate the
 Granda-Oliveros model.
This is because   if  the IR cut-off chosen is given by the particle horizon, the
Holographic dark energy  models are  not able to produce
an accelerated expansion for the present  cosmological epoch. However, if the future event horizon is used an the IR cut-off, then the  Holographic dark energy  models
have  a problem with causality. It is possible to resolve both these problems  by using Granda-Oliveros model \cite{lgo}.
 It may be noted that in the limiting case $\left\{ \alpha, \beta  \right\}
= \left\{2,1 \right\}$, $L_{GO}$ becomes proportional to the Ricci scalar curvature  $L_{GO} \propto R$.  However, based on the observational data, such a choice is
not physical. This is because the values for these parameters which best fit the observational data  have been obtained \cite{wangalfa}.
These value for   a non-flat universe   are given by  \cite{wangalfa},
\begin{eqnarray}
 \alpha  = 0.8824^{+0.2180}_{-0.1163}(1\sigma)\,^{+0.2213}_{-0.1378}(2\sigma), && \beta = 0.5016^{+0.0973}_{-0.0871}(1\sigma)\,^{+0.1247}_{-0.1102}(2\sigma).
\end{eqnarray}
These value   for a flat universe   are given by \cite{wangalfa},
\begin{eqnarray}
\alpha  = 0.8502^{+0.0984}_{-0.0875}(1\sigma)\,^{+0.1299}_{-0.1064}(2\sigma), && \beta = 0.4817^{+0.0842}_{-0.0773}(1\sigma)\,^{+0.1176}_{-0.0955}(2\sigma).
\end{eqnarray}
It may be noted that for Granda-Oliveros model, energy density $\rho_D$ can be written as
\begin{eqnarray}
\rho_D &=& 3c^2\left(\alpha H^2 + \beta \dot{H}   \right) . \label{rhoconlgo2}
\end{eqnarray}
In this paper,  we  will analyse the effect of Einstein-aether gravity on the cosmological evolution using the  Granda-Oliveros model.
We will analyse this for different models for the evolution of the scale factor of the universe, and analyse such a model for the observationally
motivated values of $\alpha$ and $\beta$.

It may be noted the  Granda-Oliveros model has been generalized to a
Chen-Jing model \cite{chens}. In this cosmological model,
the  dark energy   is a function of the Hubble parameter squared (i.e. $H^2$) and its  first and second  temporal
derivatives of the Hubble parameter $\dot{H}$ and $\ddot{H}$,
\begin{eqnarray}
\rho_D = 3c^2\left[ \alpha\left(\frac{\ddot{H}}{H}\right)   + \beta \dot{H} + \gamma H^2  \right], \label{rho}
\end{eqnarray}
where $\alpha$, $\beta$ and $\gamma$ represent three arbitrary dimensionless parameters. The inverse of the Hubble parameter, i.e. $H^{-1}$,
is introduced in the first of the three terms of Eq. (\ref{rho}), so that each of these three terms have the   dimensions.  \\
It may be noted that  in the limiting case
corresponding to $\alpha = 0$, we recover the energy density of dark energy    given by the Granda-Oliveros model
\cite{grandaoliverosa},\cite{grandaoliverosb}. Furthermore, in the limiting case,  $\alpha=0$,
$\beta=1$ and $\gamma=2$, we obtain the expression of the energy density of dark energy  with the  IR cut-off proportional to the average
radius of the Ricci scalar curvature, $L \propto R^{-1/2}$ (when  $k = 0$).
In this paper we will analyse this model for various cosmological model, with different evolution of the scale factor. We will obtain general expression
for various parameters for this dark energy model, and they can be compared to various observational data.

\section{Cosmological Models}
In this section, we will analyse the behavior of various cosmological models in Einstein-aether gravity. These will correspond to different evolution of the scale factor.
We will analyse them for the Granda-Oliveros model, using the    values obtained from observation. We will also analyse them for the Chen-Jing model.

Now we start by considering the the power-law cosmology. This cosmological model is  interesting proposal for the evolution of the scalar factor,
and it has been motivate by the existence of the  flatness and horizon  problems in the standard cosmology \cite{Rani:2014sia}.
In this cosmological model, it is possible to assume the following form for the evolution
      of the scale factor,
\begin{eqnarray}
a\left( t \right) = a_0 t^m. \label{21}
\end{eqnarray}
where $a_0$ is the present day value of $a\left(t\right)$.  It is also important to only consider  $m>0$, and this will produce an accelerating universe \cite{Rani:2014sia}.
It may be noted that  for $m > 1$, the power-law cosmology
can solve the horizon problem, the  flatness problem, and the    problem associated with age of  the early universe \cite{power1,power11}.
 The power-law cosmology has been used for analysing the cosmological behavior in     modified theories of gravity  \cite{power2,power22}.
 Now for Einstein-aether gravity, it is possible to analyse the Granda-Oliveros cut-off for power-law cosmological model. Thus,   for a
 non-flat universe, we obtain,
\begin{eqnarray}
F\left(K\right) &=& \frac{2 c^2 K ( 0.8824m  -0.5016 )}{ \varepsilon  m}+C_1\sqrt{K}.
\end{eqnarray}
For a flat universe, we obtain,
\begin{eqnarray}
F\left(K\right) &=& \frac{2 c^2 K ( 0.8502m  -0.4817 )}{ \varepsilon  m}+C_1\sqrt{K}.
\end{eqnarray}
It is also possible to analyse the power-law cosmology using the Chen-Jing model,  we have
\begin{eqnarray}
F\left( K\right) &=& \frac{2 c^2 K \left[2 \alpha +n (n \beta +\gamma )\right]}{ \varepsilon  n^2}+C_4\sqrt{K}.
\end{eqnarray}
In this section,  $C_i$ will denote integration constants, for example,  here $C_1, C_4$ denotes the integration constants.

It is also possible to consider a different kind of power law \cite{a, za},
\begin{eqnarray}
a\left( t \right) = a_0 (t_s  -t )^{-n} \label{37},
\end{eqnarray}
where $n>0$ and $t<t_s$. Such models have a  future singularity at finite time, and this is denoted by  $t_s$.
The   Granda-Oliveros cut-off   for such model can be analysed for both a flat universe and a non-flat universe.
For a non-flat universe, we obtain
\begin{eqnarray}
  F\left( K\right) &=&  \frac{2 c^2 K (0.8824n +0.5016 )}{ \varepsilon  n}+C_3\sqrt{K}.
 \end{eqnarray}
For a flat universe, we obtain
\begin{eqnarray}
 F\left( K\right) &=&  \frac{2 c^2 K (0.8502n +0.4817 )}{ \varepsilon  n}+C_3\sqrt{K}.
\end{eqnarray}
Now using Chen-Jing model, we obtain
\begin{eqnarray}
 F\left( K\right) =\frac{2 c^2 K \left[2 \alpha +n (n \beta +\gamma )\right]}{ \varepsilon  n^2}+C_4\sqrt{K},
\end{eqnarray}

It is also possible to analyse intermediate inflation in Einstein-aether gravity. The  intermediate inflation has been used
to obtain  exact analytic solutions for a given class of potentials for the inflation. The scale factor for intermediate inflation
can be expressed as   \cite{18, 31},
\begin{eqnarray}
a\left( t \right) =  e^{B t^{\theta }}, \label{69}
\end{eqnarray}
where $B >0$ and $0< \theta <1$.
For   a non-flat universe, we have
\begin{eqnarray}
F\left( K \right) &=&  2(B \varepsilon  \theta )^{-1} c^2 K    \left(\frac{K M^2}{3B^2 \varepsilon  \theta ^2}\right)^{\frac{-\theta}{2(-1+\theta )}}
\nonumber \\ && \times \left[-0.5016 (-1+\theta )^2+0.8824B \theta
\left(\frac{K M^2}{3B^2 \varepsilon  \theta^2}\right)^{\frac{\theta}{2 (-1+\theta )}}  \right] \nonumber \\
&&+\sqrt{K}C_7.
\end{eqnarray}
For a flat universe, we have
\begin{eqnarray}
 F\left( K \right) &=&  2(B \varepsilon  \theta )^{-1} c^2 K    \left(\frac{K M^2}{3B^2 \varepsilon  \theta ^2}\right)^{\frac{-\theta}{2(-1+\theta )}}\nonumber \\&&\times
 \left[-0.4817 (-1+\theta )^2+0.8502 \theta
 \left(\frac{K M^2}{3B^2 \varepsilon  \theta^2}\right)^{\frac{\theta}{2 (-1+\theta )}}  \right] \nonumber \\
&&+\sqrt{K}C_7.
\end{eqnarray}
Now for Chen-Jing model, we obtain
\begin{eqnarray}
F\left( K \right) &=&\varepsilon^{-1} 2\ 3^{-4^{\theta } \left(\frac{1}{-1+\theta }\right)^{-2 \theta }-2^{1-\theta } \left(\frac{1}{-1+\theta
}\right)^{\theta }} c^2 K \left(\frac{K M^2}{B^2 \varepsilon  \theta ^2}\right)^{4^{\theta } \left(\frac{1}{-1+\theta }\right)^{-2 \theta }}\nonumber \\
&&\times \left\{\frac{2^{\theta} \beta  \left(\frac{1}{-1+\theta }\right)^{2 \theta } \left(\frac{K M^2}{B^2 \varepsilon  \theta ^2}\right)^{2^{1-\theta } \left(\frac{1}{-1+\theta}\right)^{\theta }}}{2^{1+3 \theta }+2^{\theta } \left(\frac{1}{-1+\theta }\right)^{2 \theta }+4 \left(\frac{1}{-1+\theta }\right)^{3 \theta }} \right. \nonumber \\
&&\left.
+\frac{2^{\theta} 3^{2^{-\theta } \left(\frac{1}{-1+\theta }\right)^{\theta }} \gamma  \left(\frac{1}{-1+\theta }\right)^{-1+2 \theta } \left(\frac{K M^2}{B^2 \varepsilon \theta ^2}\right)^{2^{-\theta } \left(\frac{1}{-1+\theta }\right)^{\theta }}}{2^{1+3 \theta } B \theta +2^{\theta } B \left(\frac{1}{-1+\theta }\right)^{2 \theta } \theta +2 B \left(\frac{1}{-1+\theta }\right)^{3 \theta } \theta } \right. \nonumber \\
&&\left.+\frac{3^{2^{1-\theta } \left(\frac{1}{-1+\theta }\right)^{\theta }} \alpha  \left(2-3 \theta +\theta ^2\right)}{B^2 \left[1+2^{1+2 \theta } \left(\frac{1}{-1+\theta }\right)^{-2 \theta }\right] \theta ^2}\right\},
 \\
&& + C_8\sqrt{K}.
\end{eqnarray}

It is possible to analyse universes with no past time-like singularity. In such cosmological models, the  universe in the infinite past there exists a
static state of cosmology, and   this state evolves
into an inflationary stage. The scale factor for such  cosmological models can be expressed as  \cite{17, 30},
\begin{eqnarray}
a\left( t \right) =  A \left(B+e^{n t}\right)^{\lambda }
\end{eqnarray}
where $A$, $B$, $n$ and $\lambda$ are four positive constant parameters. In  order to avoid singularities, we have to use  $B>0$. Furthermore, for
for the positivity of scale factor, we  have to use $A>0$. It may be noted that in this model, for
$a < 0$ or $\lambda <0$,   a  singularity exists. So, for
 the expanding model, we have to only consider  $a >0$ and  $\lambda >0$.
Using  the Granda-Oliveros cut-off for this  model, we can analyse a flat universe and a non-flat universe.
So, for a non-flat universe, we have
\begin{eqnarray}
  F\left(K\right) &=&\frac{c^2 K \left(1.7648 \lambda -1.0032 + 0.5016  \sqrt{\frac{3\varepsilon n^2 \lambda ^2}{K M^2}} \log K\right)}{\varepsilon  \lambda}\nonumber \\
  && + \sqrt{K} C_5.
 \end{eqnarray}
For a flat universe, we have
\begin{eqnarray}
  F\left(K\right) &=& \frac{c^2 K \left(1.7004 \lambda -0.9634 + 0.4817  \sqrt{\frac{3\varepsilon n^2 \lambda ^2}{K M^2}} \log K\right)}{\varepsilon  \lambda}\nonumber \\
  && + \sqrt{K} C_5.
 \end{eqnarray}
We can also use the Chen-Jing model, and obtain
\begin{eqnarray}
F\left(K\right) &=& c^2 ( \varepsilon M^2 \lambda ^2)^{-1} \left[-6 \varepsilon n^2 \alpha  \lambda ^2+2 K M^2 \left(2 \alpha -\gamma  \lambda+\beta  \lambda ^2\right)
\right. \nonumber \\
&&\left. +\sqrt{3} K M^2 \sqrt{\frac{\varepsilon n^2 \lambda ^2}{K M^2}} \left(-3 \alpha +\gamma  \lambda \right) \log K\right], \nonumber \\
&&+ C_6\sqrt{K}.
\end{eqnarray}

It is possible to analyse   matter dominated universe  and the accelerated phase of the universe using a single formalism \cite{md, dm}.
In such cosmological models, the
 Hubble constant  is given by \cite{hep-th/0506212, ycyc}
\begin{equation}\label{hubbleassume}
H(t)=H_0+\frac{H_1}{t},
\end{equation}
where $H_0$ and $H_1$ are two constant parameters.
In this case, for the non-flat universe, we obtain
\begin{eqnarray}
F\left(K\right) &=&  - 2 c^2 (9 \varepsilon M^2 H_1)^{-1} [-9.0288 \varepsilon^2 K M^2   H_0- 13.5431\varepsilon   H_0^2
\nonumber \\ && +K M^2(0.5016\varepsilon^3 K M^2  -7.9416  H_1)
\nonumber \\
&&+\sqrt{K} C_{10}.
\end{eqnarray}
Furthermore, for the flat universe, we obtain
\begin{eqnarray}
  F\left(K\right) &=& - 2 c^2 (9 \varepsilon M^2 H_1)^{-1} [-8.6706 \varepsilon^2 K M^2
 H_0-13.0059 \varepsilon  H_0^2\nonumber \\ && +K M^2  (0.4817\varepsilon^3 K M^2  -7.5618  H_1 ) ]  \nonumber \\
&&+\sqrt{K} C_{10}.
 \end{eqnarray}
Now using the Chen-Jing model, we obtain
\begin{eqnarray}F(K)&=&\sqrt{K}\left[C_{11}+\,{\frac {2{c}^{2}{M}^{2}\alpha\,{K}^{3/2}}{\epsilon\,{H_{{1}}}^{2}}}-
{\frac {{c}^{2}{M}^{2}\beta\,{K}^{3/2}}{\epsilon\,H_{{1}}}}\right. \nonumber \\
&&\left.+{\frac {{c
}^{2}{M}^{2}\gamma\,{K}^{3/2}}{\epsilon}} -\,{\frac {9{c}^{2}\sqrt {2}M
H_{{0}}\alpha\,K}{\sqrt {\epsilon}{H_{{1}}}^{2}}} +\,{\frac {3{c}^{2}
\sqrt {2}MH_{{0}}\beta\,K}{\sqrt {\epsilon}H_{{1}}}}\right. \nonumber \\
&&\left.+\,{\frac {
18\sqrt {2\epsilon}{c}^{2}{H_{{0}}}^{2}H\alpha\,\ln  \left( K
 \right) }{{H_{{1}}}^{2}M}}-\,{\frac {3\sqrt {2\epsilon}{c}^{2}{H_{{0}}}^{2}H\beta\,\ln  \left( K \right) }{H_{{1}}M}} \right.\nonumber \\
&&\left.-\,{\frac {
6\sqrt {2\epsilon}{c}^{2}\alpha\,{H_{{0}}}^{3}\ln  \left( K
 \right) }{{H_{{1}}}^{2}M}}
\right]. \label{maurobiglino1}
\nonumber \\
\end{eqnarray}
The quantum deformed de Sitter ($q$-de Sitter) solution has been obtained by
a quantum deformation of the quantum deformation of the conformal group \cite{qd}. In fact,
the $q$-deformed de Sitter solution has also been used in the analyzing of dS/CFT correspondence and the entanglement entropy
for such a solution has also been obtained \cite{qd}. The $q$-de Sitter has also been used for analyzing cosmology \cite{Setare:2014vna}.
Now we will analyse this model for the  $q$-de Sitter scale factor
\cite{Setare:2014vna},
\begin{eqnarray}
&&a(t)=e_{q}(H_0 t)=\Big[1+(q-1)H_0 t\Big]^{\frac{1}{q-1}}\label{a(t)}.
\end{eqnarray}
In this model it is possible to  interpolate between the cosmological model based on a  power-law and the cosmological model involving
de Sitter spacetime.
In fact, for  early times, $H_0 t\gg 1$, we obtain
\begin{eqnarray}
a_{early}(t)\sim \Big[H_0 t\Big]^{\frac{1}{q-1}}=t^{p}.
\end{eqnarray}
It is possible to have an  acceleration expansion, when $p>1$, and  $q<2$. So,  we can write
\begin{eqnarray}
a_{early}(t)\preceq e_{q}(H_0 t)\preceq a_{dS}(t)\label{inequality}.
\end{eqnarray}
This inequality,  given in Eq. (\ref{inequality}), produces interesting  cosmological evolution in  $q$-de Sitter.
The $q$-de Sitter  can be used to smoothly connect  the  early cosmological  epoch to late time  evolution universe.
Now we can analyse such a model for a non-flat universe as
\begin{eqnarray}
  F(K) &=& 6 c^2 H_0^2 \left(\frac{K}{\varepsilon H_0^2}\right)^{1+\frac{1}{-2+q}} \left[\left(\frac{K}{\varepsilon H_0^2}\right)^{\frac{1}{-2+\frac{2}{-1+q}}}\right]
 ^{\frac{1}{-1+q}} \nonumber \\&&\times   \left[e^{\frac{q \left((-2+q) \log[K]+\log\left[\frac{K}{\varepsilon H_0^2}\right]+2 (-2+q) \log\left[\left(\frac{K}{\varepsilon
H_0^2}\right)^{-\frac{-1+q}{2 (-2+q)}}\right]\right)}{2 (-2+q)^2 (-1+q)}}\right.\nonumber \\
 && \left. \times   K^{-\frac{q}{4-6 q+2 q^2}}\left(\frac{K}{\varepsilon H_0^2}\right)^{-\frac{-1+2q}{2 (-2+q)^2 (-1+q)}} 0.8824 \right.\nonumber \\
 && \left. \times
 -\frac{(-2+q)^2 \cdot 0.5016 }{-1+q}\right]    M^{-2} + \sqrt{K}C_{11}.
\end{eqnarray}
We can also analyse such a model for the flat universe as
\begin{eqnarray}
 F(K) &=& 6 c^2 H_0^2 \left(\frac{K}{\varepsilon H_0^2}\right)^{1+\frac{1}{-2+q}} \left[\left(\frac{K}{\varepsilon H_0^2}\right)^{\frac{1}{-2+\frac{2}{-1+q}}}
 \right]^{\frac{1}{-1+q}} \nonumber \\&& \times \left[e^{\frac{q \left((-2+q) \log[K]+\log\left[\frac{K}{\varepsilon H_0^2}\right]+2 (-2+q)
 \log\left[\left(\frac{K}{\varepsilon
H_0^2}\right)^{-\frac{-1+q}{2 (-2+q)}}\right]\right)}{2 (-2+q)^2 (-1+q)}} \right.\nonumber \\
 && \left. \times  K^{-\frac{q}{4-6 q+2 q^2}} \left(\frac{K}{\varepsilon H_0^2}\right)^{-\frac{-1+2q}{2 (-2+q)^2 (-1+q)}} 0.8502
 \right.\nonumber \\
 && \left. \times -\frac{(-2+q)^2 \cdot 0.4817 }{-1+q}\right]   M^{-2} + \sqrt{K}C_{11}.
\end{eqnarray}
We can use the Chen-Jing model, and obtain
\begin{eqnarray}
F(K) &=& \sqrt {K} C_{13}+\,{\frac {3{c}^{2} \left( 2\,
\alpha\,{q}^{2}-4\,\alpha\,q+2\,\alpha-\beta\,q+\beta+\gamma \right) }
{\epsilon}}K. \nonumber \\
\end{eqnarray}
Thus, we can see the    Einstein-aether gravity modifies the cosmological evolution in various different model, with different evolution of scale factor.
Hence, this deformation is almost a universal feature of  Einstein-aether gravity.
We have calculated $F(K)$ for these different cosmological models. Thus, these cosmological model directly depend on the aether vector field.
It is also possible to calculate   $L_{GO} $, $\omega_{EA}$, $\rho_{EA}$ and $p_{EA}$ for these various
different cosmological models. It can be argued that these quantities will also
depend on the aether field (see Appendix).  Hence, the breaking of Lorentz symmetry by the introduction of a time-like aether vector field can modify the
cosmological dynamics in a non-trivial way. Here we explicitly calculated such a modification for a large number of cosmological models.

\section{\emph{Om} Diagnostic  Analysis  }
\begin{figure} \centering
\includegraphics[width=8cm]{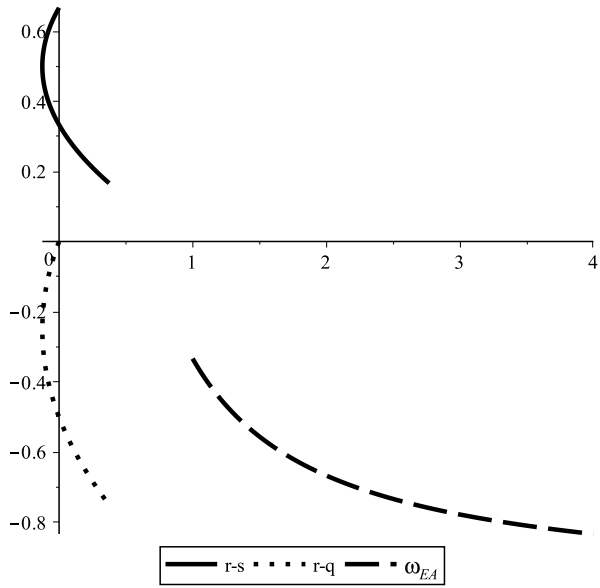}
 \caption{ $r-s$, $r-q$, $m-\omega_{Einstein-aether  }$ for power-law  in the redshift $t=(1+z)^{-1/m}$ range $0.07\leq z \leq 2.3 $.}\label{rsqw-power}
\end{figure}

\begin{figure} \centering
\includegraphics[width=8cm]{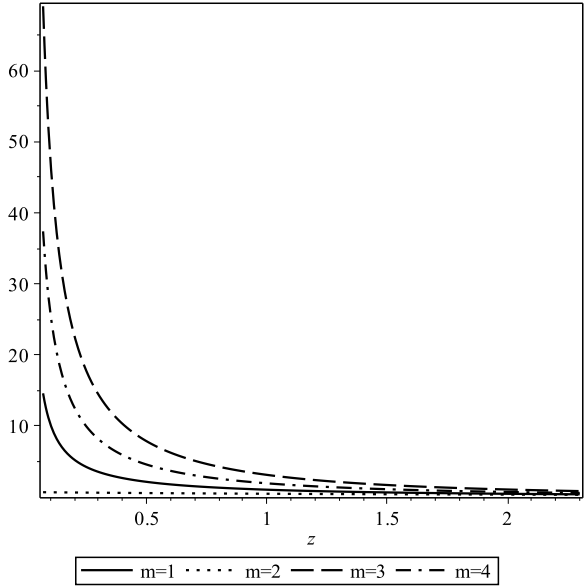}
 \caption{ $Om(z)$ for power-law   in the redshift $t=(1+z)^{-1/m}$  range $0.07\leq z \leq 2.3 $.}\label{Om(z)-power}
\end{figure}

In this section, we will perform the \emph{Om} diagnostic  analysis of various different cosmological models.
The  cosmological parameters like the Hubble parameter $H$,  deceleration parameter  $q$,    and the Equation of State (EoS) parameter
$\omega$ are important to understand the behavior of  cosmological models. It is theoretically and observationally known that  different dark energy
models produce  a positive Hubble parameter
and a negative deceleration parameter
(i.e. $H > 0$ and $q < 0$, for  the present cosmological epoch. So,
$H$ and $q$ can not be used to  effectively differentiate between  the different dark energy   models. Therefore, a higher order of time derivatives of $a(t)$ is required to    analyse  the dark energy  models  \cite{alam, sah}. So,  third order temporal  derivative of $a(t)$ can be used to resolve the problem that most dark energy models produce   $H > 0$ and $q < 0$ for  the present cosmological epoch.
  So, now the  statefinder parameters   $\left\{r,s\right\}$, can be expressed as
\begin{eqnarray}
r &=& \frac{\dddot{a}}{aH^3}, \label{r1}\\
s &=&   \frac{r -1}{3\left(q-1/2\right)}, \label{s1}
\end{eqnarray}
where $q$ represents the deceleration parameter, which is given by
\begin{eqnarray}
q=-\frac{1}{H^2}\frac{\ddot{a}}{a}.
\end{eqnarray}
An alternative way to write   $r$ and $s$ is as
\begin{eqnarray}
r&=& 1 + 3\frac{\dot{H}}{H^2}+ \frac{\ddot{H}}{H^3}, \label{r2}\\
s&=& -\frac{3H\dot{H}+\ddot{H}}{3H\left( 2\dot{H}+3H^2  \right)} \nonumber \\
&=&  -\frac{3\dot{H}+\ddot{H}/H}{3\left( 2\dot{H}+3H^2  \right)}. \label{s2}
\end{eqnarray}
It may be noted that statefinder parameters  $\left\{r, s\right\} = \left\{1, 0\right\}$ represents  the point where  the flat $\Lambda$CDM model  exists in the $r-s$ plane \cite{huang}.
So, the departure of dark energy  models from this fixed point can be  used to obtain  the distance of these models from the flat $\Lambda$CDM model,
taken as reference model. \\
We also note that  in the $\left \{r, s\right \}$ plane, a positive value of the parameter $s$ (i.e. $s > 0$) indicates a quintessence-like model of dark energy, and a negative value of the parameter $s$ (i.e. $s < 0$) indicates a phantom-like model of dark energy . Furthermore,  the evolution from phantom to quintessence is obtained by crossing of the point $\left\{r, s\right\} = \left\{1, 0\right\}$ in the $\left \{r,s \right \}$ plane \cite{wu1}.\\
So, different cosmological   models, like the models with a cosmological constant $\Lambda$, braneworld models , chaplygin gas and quintessence models,
have been studied using such an analysis  \cite{alam}. In this study, it was argued that $\left\{r,s\right\}$
can be used to  differentiate between different  models. An analysis based on  $\left\{r,s\right\}$ has also been used to  differentiate between dark energy
and modified gravity   \cite{wang, wu1}. \\
An important geometrical diagnostic which can be used to for such analysis is called the    \emph{Om} diagnostic analysis \cite{Sahni}.
Usually in the study of the  statefinder parameters $r$ and $s$,   higher order temporal  derivatives of $a(t)$ are used. However, in the  \emph{Om} diagnostic analysis
only
first order temporal  derivative are used. This is because it only involves the Hubble parameter,
and the   Hubble parameter depends on a single time derivative of $a(t)$.
So,   the \emph{Om} diagnosis  can be considered as a simpler diagnostic than the statefinder diagnosis  \cite{Shahalam:2015lra}.
It may be noted that the  \emph{Om} diagnosis has also been applied to   Galileons models
\cite{Jamil:2013yc,deFromont:2013iwa}. This set of parameters can now be represented as
\begin{eqnarray}
Om(z)=\frac{\left[\frac{H(z)}{H_0}\right]^2-1}{(1+z)^3-1}.
\end{eqnarray}
For a constant EoS parameter $\omega$,  the expression for $Om(z)$ is given by
\begin{eqnarray}
&&Om(z)= \Omega_{m0} + (1 -\Omega_{m0})\frac{(1+z)^{3(1+\omega)}-1}{(1+z)^3-1},
\end{eqnarray}
Thus,  we observe   that we have different values of  $Om(z) =\Omega_{m0}$  for the  $\Lambda$CDM model,
 quintessence, and phantom cosmological models.
In Figure \ref{rsqw-power}, we plot the first cosmological parameters $r-s$, $r-q$, $m-\omega_{Einstein-aether  }$ for power-law
in the redshift $t=(1+z)^{-1/m}$ range $0.07\leq z \leq 2.3 $. A continuous behavior is observed.
For $r-s$, we observe that when $r$ is increasing, $s$ is starting to decreasing monotonically, never vanishes.
A similar pattern is repeating but in the negative range for $q$. As we observe, the  observational value for $q\approx -0.67$
exists in this model. In Figure \ref{Om(z)-power}, we plot $Om(z)$  for power-law  in the redshift $t=(1+z)^{-1/m}$ range $0.07\leq z \leq 2.3 $.
We   observe    that when the redshift $z$ is increasing within the interval  $0.07\leq z \leq 2.3 $, the $Om(z)$ is decreasing monotonically.
For all power exponents $m\neq2$, always $Om(z)>\Omega_{m0}$, so the model mimics a quintessence with effective EoS $w>-1$.

\begin{figure} \centering
\includegraphics[width=8cm]{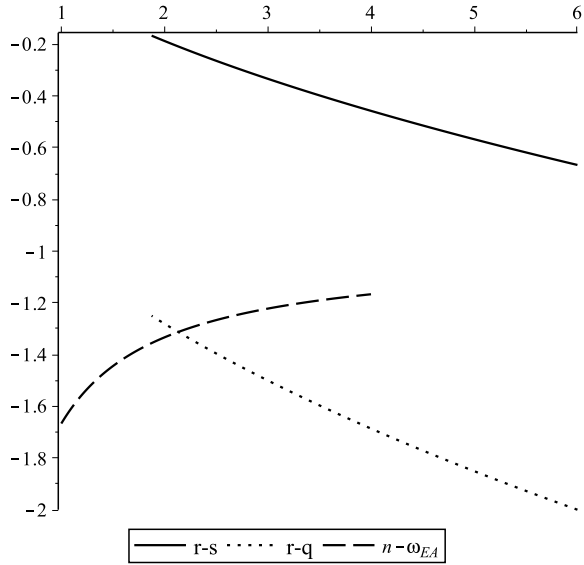}
 \caption{ r-s, r-q, $\omega_{Einstein-aether  }$ for model with future singularity in the redshift $t=t_s-(1+z)^{1/n} $ range $0.07\leq z \leq 2.3 $.}\label{rsqw-future}
\end{figure}

\begin{figure}[hc] \centering
\includegraphics[width=8cm]{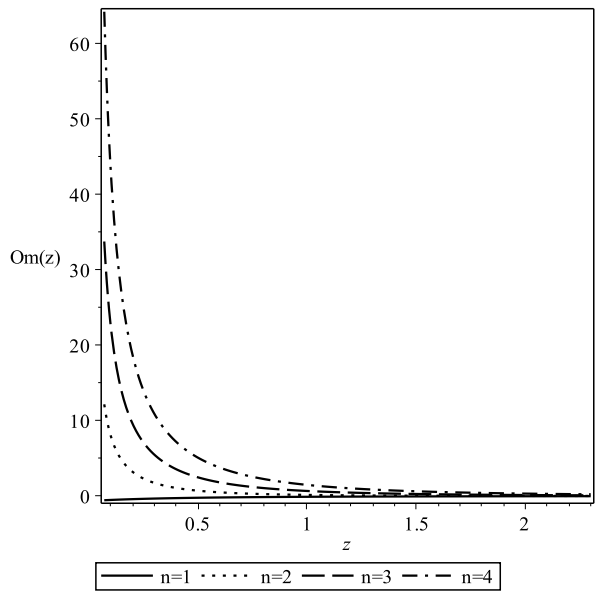}
 \caption{ $Om(z)$ for model with  future singularity   in the redshift $t=t_s-(1+z)^{1/n}$ range $0.07\leq z \leq 2.3 $.}\label{Om(z)-future}
\end{figure}

In Figure \ref{rsqw-future}, we plot the first cosmological parameters $r-s$, $r-q$, $\omega_{Einstein-aether  }$
for a cosmological  model with a future singularity. In the redshift $t=t_s-(1+z)^{1/n} $ range $0.07\leq z \leq 2.3 $, a
 continuous behavior is observed. For $r-s$, we observe that when $r$ is increasing, $s$ is starting to decreasing monotonically,
 always remains negative. A similar pattern is repeating but in the negative ranges for $r,q$. As we see, the   observational value for $q\approx -0.67$
 do not exist in this  model. In Figure \ref{Om(z)-future}, we plot $Om(z)$  for future singularities model in the redshift $t=t_s-(1+z)^{1/n} $
 range $0.07\leq z \leq 2.3 $.  We observe  that when the redshift $z$ is increasing within the interval  $0.07\leq z \leq 2.3 $, the $Om(z)$ is decreasing
 monotonically, always $Om(z)>\Omega_{m0}$. So, the model with future singularity  mimics a quintessence with effective EoS $w>-1$.

\begin{figure}[hc] \centering
\includegraphics[width=8cm]{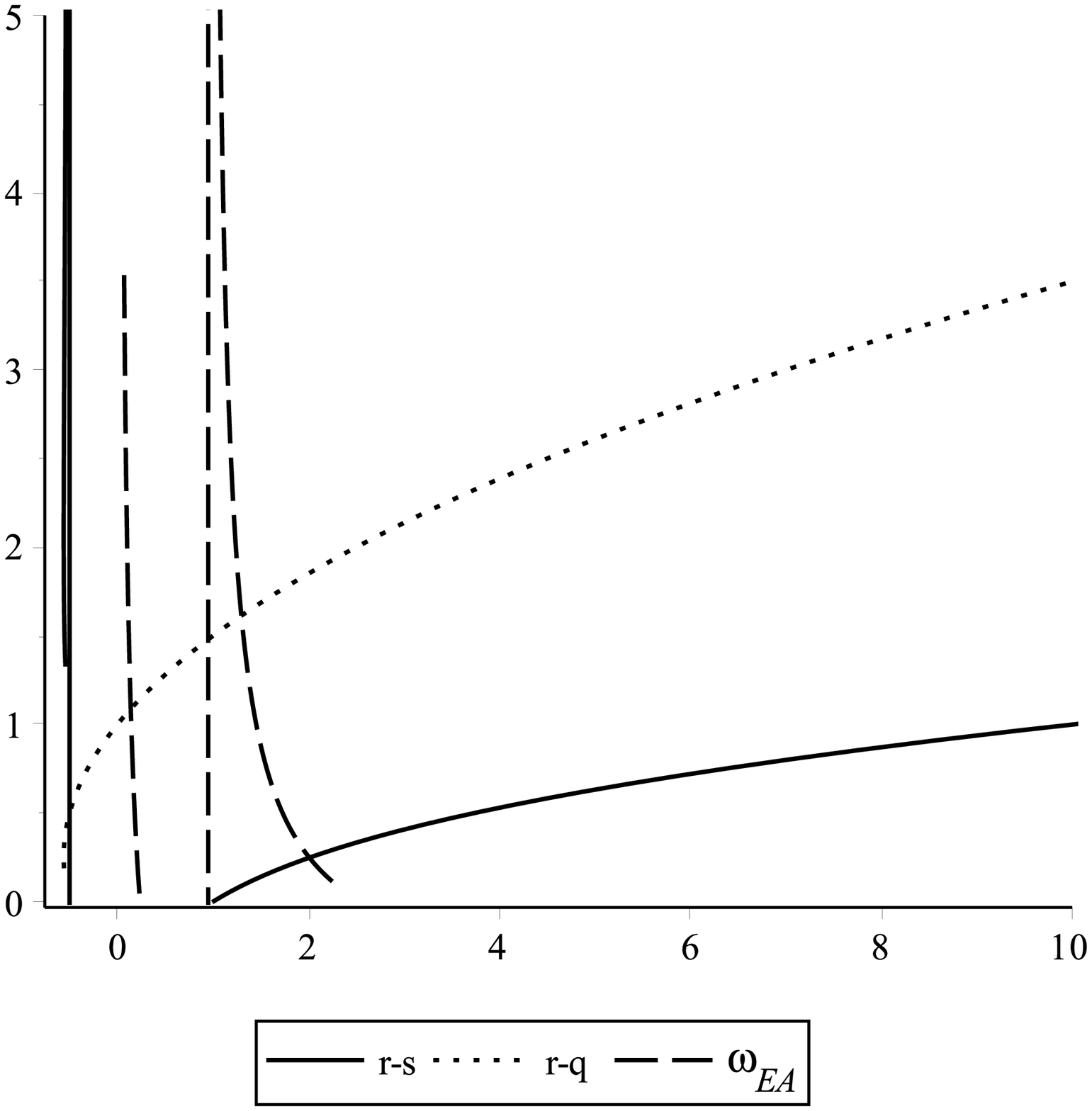}
 \caption{ r-s, r-q, $\omega_{Einstein-aether  }$ for emergent universe  in the redshift  $t=\ln  \left( {{\rm e}^{-{\frac {\ln  \left( A+zA \right) }{\lambda}}}
}-B \right) {n}^{-1}$  range $0.07\leq z \leq 2.3 $ with parameters $A = 1; B = 1; \lambda= 2; n = 1$.}\label{rsqw-emergent}
\end{figure}

\begin{figure}[hc] \centering
\includegraphics[width=8cm]{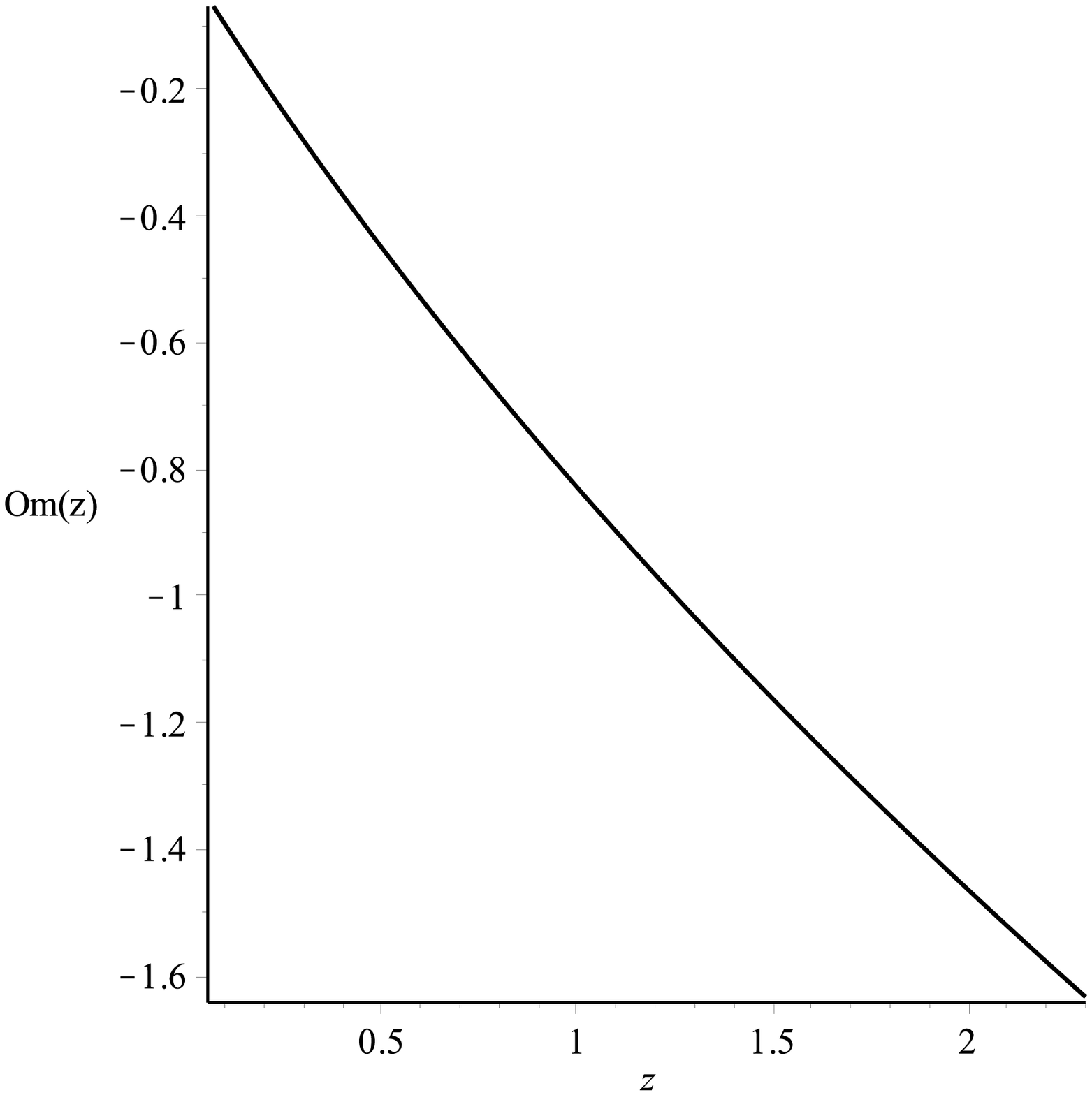}
 \caption{$Om(z)$ for emergent  universe  in the redshift $t=\ln  \left( {{\rm e}^{-{\frac {\ln  \left( A+zA \right) }{\lambda}}}
}-B \right) {n}^{-1}$ range $0.07\leq z \leq 2.3 $ with parameters $A = 1; B = 1; \lambda= 2; n = 1$.}\label{Om(z)-emergent}
\end{figure}

In Figure \ref{rsqw-emergent}, we plot the first cosmological parameters $r-s$, $r-q$, $\omega_{Einstein-aether  }$  for models of emergent universe,
in the redshift  $t=\ln  \left( {{\rm e}^{-{\frac {\ln  \left( A+zA \right) }{\lambda}}}
}-B \right) {n}^{-1}$ range $0.07\leq z \leq 2.3 $. A continuous behavior is observed. For $r-s$, we observe that when $r$ is increasing, $s$   starts to
increase monotonically too, always remaining  positive. A similar pattern is repeating  for $r,q$. As we observe, the   observational value for $q\approx -0.67$
do not  exist in this  model. In Figure \ref{Om(z)-emergent}, we plot $Om(z)$  for emergent universe  in the redshift  $t=\ln  \left( {{\rm e}^{-{\frac{\ln
\left( A+zA \right) }{\lambda}}}
}-B \right) {n}^{-1}$  range $0.07\leq z \leq 2.3 $. We observe that when the redshift $z$ is increasing within the interval  $0.07\leq z \leq
2.3 $, the $Om(z)$ is  monotonically decreasing, and  $Om(z)<\Omega_{m0}$, so this model mimics a phantom model with effective EoS $w<-1$.

\begin{figure}[hc] \centering
\includegraphics[width=8cm]{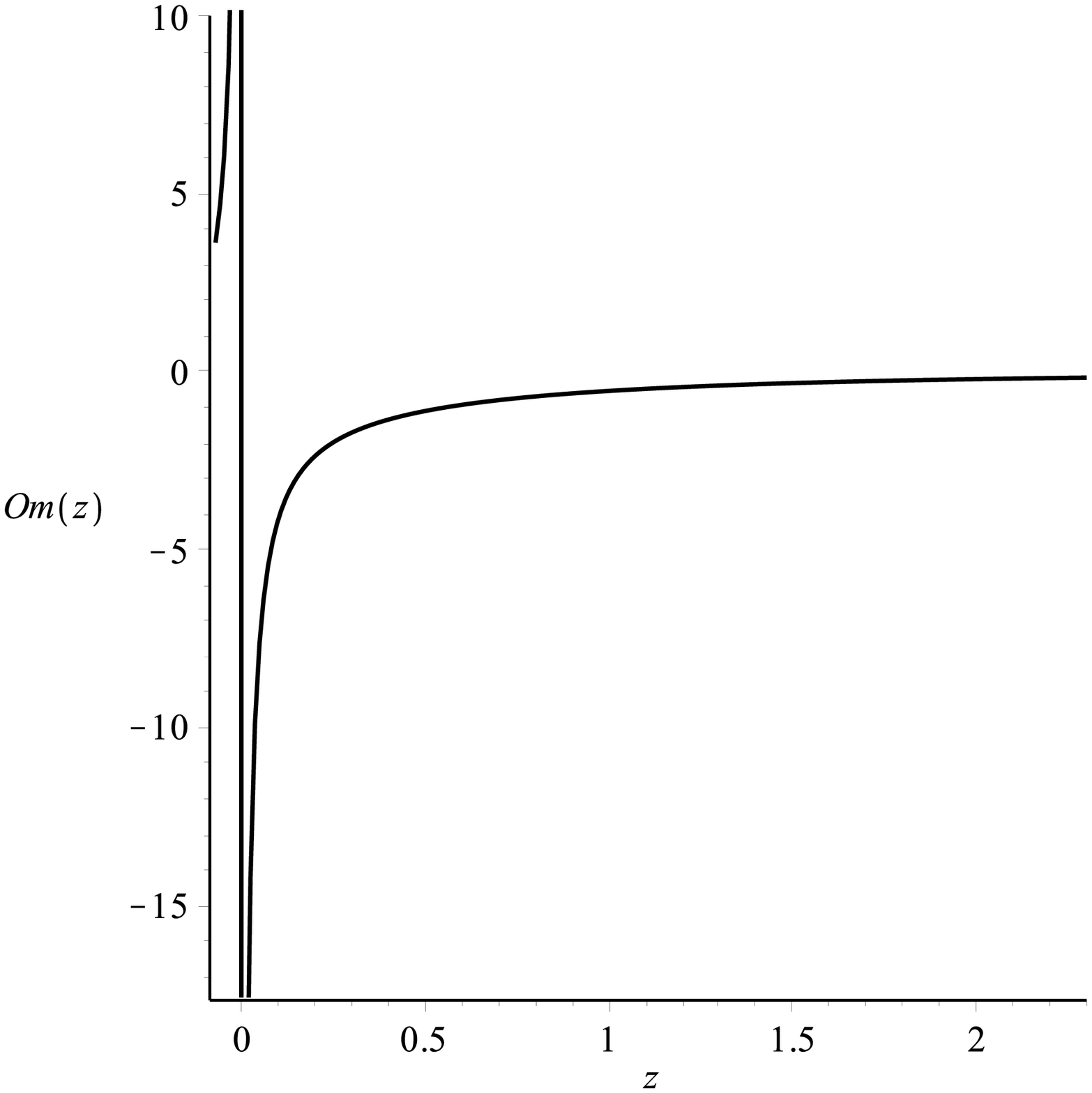}
 \caption{ $Om(z)$ for intermediate inflation    in the redshift $t=\Big[-\frac{\ln(1+z)}{B}\Big]^{\theta^{-1}}$ range $0.07\leq z \leq 2.3 $ with parameters $\theta= 2; B = 1$.}\label{rsqw-intermediate}\label{Om(z)-intermediate}
\end{figure}

In Figure \ref{rsqw-intermediate}, we plot the first cosmological parameters $r-s$, $r-q$, $\omega_{Einstein-aether  }$  for  for intermediate inflation,
in the redshift $t=\Big[-\frac{\ln(1+z)}{B}\Big]^{\theta^{-1}}$ range $0.07\leq z \leq 2.3 $.
A continuous behavior is observed. For $r-s$, we observe that when $r$ is increasing, $s=-1$ remains constant.
For $r,q$ when we are increasing $r$, $q$ is decreasing and $-1<q<0$. As we observe, the approved observational value for
$q\approx -0.67$ do not exist in  this model. In Figure \ref{Om(z)-intermediate},  we plot $Om(z)$  for intermediate inflation
in the redshift $t=\Big[-\frac{\ln(1+z)}{B}\Big]^{\theta^{-1}}$  range $0.07\leq z \leq 2.3 $. We observe that when the redshift $z$ is increasing within the interval
$0.07\leq z \leq 2.3 $, the $Om(z)$ is monotonically increasing, and  $Om(z)<\Omega_{m0}$, so the intermediate  inflation  mimics a phantom model with effective EoS $w<-1$.

\begin{figure}[hc] \centering
\includegraphics[width=8cm]{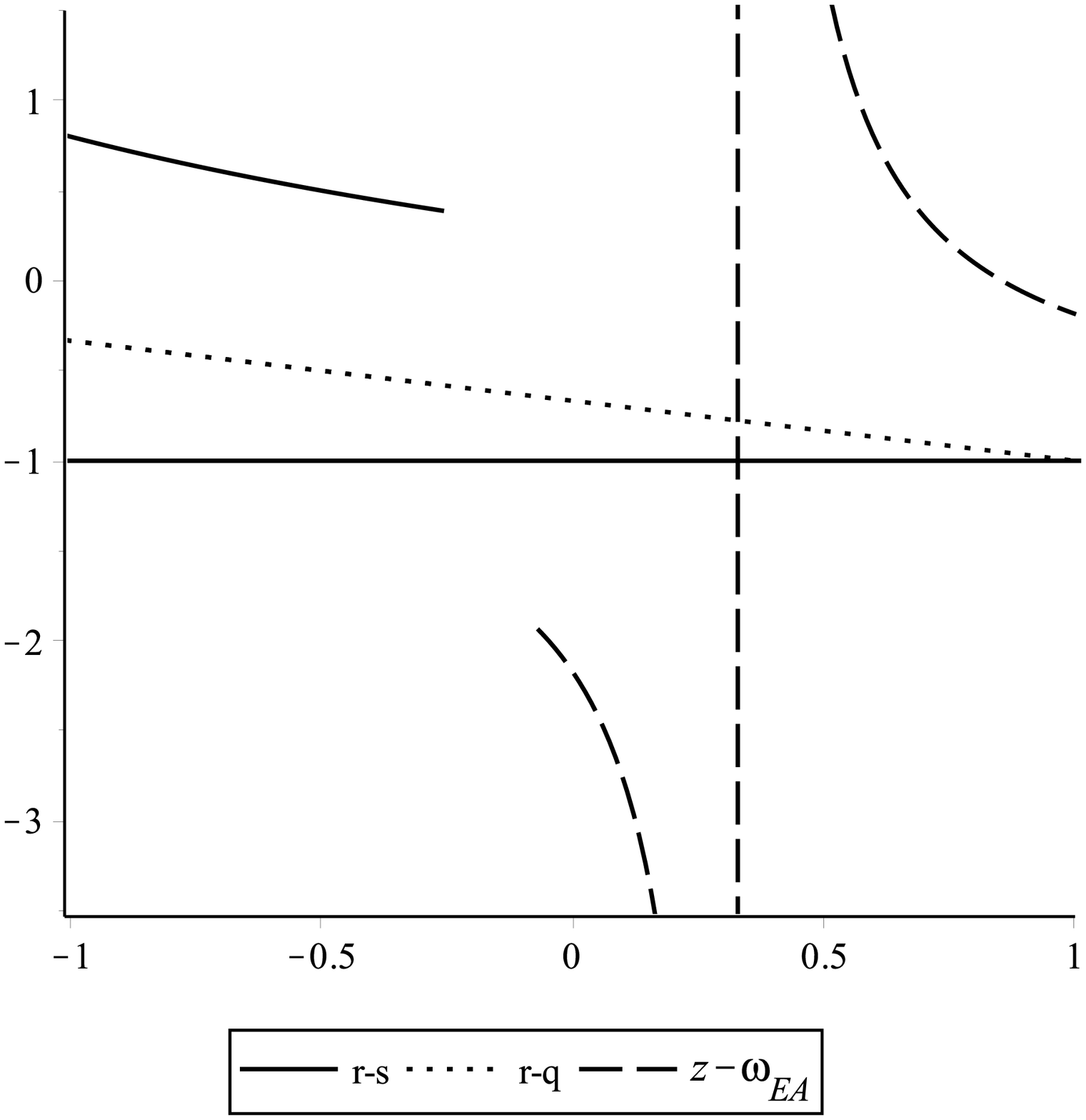}
 \caption{ r-s, r-q, $\omega_{Einstein-aether  }$ for intermediate inflation   in the redshift $t=\Big[-\frac{\ln(1+z)}{B}\Big]^{\theta^{-1}}$ range $0.07\leq z \leq 2.3 $ with parameters $\theta= 2; B = 1$.}\label{rsqw-intermediate}
\end{figure}

\section{Observational Constraints}
In this section, we will apply observational data from Ia Supernovae Ia, baryonic acoustic oscillations (BAO),
and data of the Hubble parameter $H$ to study the constraints on parameters of different cosmological models.
The total $\chi^2$ for joint data set which we use is defined by
\begin{align}
\chi_{\rm tot}^2=\chi_{\rm SN}^2+\chi_{\rm BAO}^2+\chi_{\rm Hub}^2 \,,
\label{chi_tot}
\end{align}
where the $\chi^2_i$ for each set of data is evaluated. To compute it we need    the  luminosity distance
 $D_{L}(z)$.

\begin{table}
\caption{Values of $\frac{d_A(z_\star)}{D_V(Z_{BAO})}$ for distinct values of $z_{BAO}$ \cite{Farooq:2013hq}.}
\begin{center}
\label{hubble}
\begin{tabular}{cc}
\hline\hline
~$z_{BAO}$ & ~~ $\frac{d_A(z_\star)}{D_V(Z_{BAO})}$ \\
0.106&~~    $30.95 \pm 1.46$ \\
0.2&~~    $17.55 \pm 0.60$\\
0.35&~~   $10.11 \pm 0.37$\\
 0.44&~~   $8.44 \pm 0.67$\\
 0.6&~~    $6.69 \pm 0.33$\\
0.73&~~    $5.45 \pm 0.31$\\

\hline\hline
\end{tabular}
\end{center}
\end{table}

The luminosity distance is defined as
 \begin{equation}
D_L(z)=(1+z)
\int_0^z\frac{H_0dz'}{H(z')}.
\end{equation}
We  use the distance modulus $\mu$,   which is given by
\begin{eqnarray}
  \mu=m - M=5 \log D_L+\mu_0,
\end{eqnarray}
 where $m$ and $M$ are defined as the apparent and absolute magnitudes of the Supernovae. Here
$\mu_0=5 \log\left(\frac{H_0^{-1}}{\rm Mpc}\right)+2 5$ is a
nuisance parameter (which will be marginalized). We then have that the corresponding $\chi^2$
for this data set,
\begin{align}
\chi_{\rm SN}^2(\mu_0,\theta)=\sum_{i=1}^{580}
\frac{\left[\mu_{th}(z_i,\mu_0,\theta)-\mu_{obs}(z_i)\right]^2}{
\sigma_\mu(z_i)^2}\,,
\end{align}
where $\mu_{obs}$, $\mu_{th}$ and $\sigma_{\mu}$  indicates the observed distance modulus, the theoretical distance
modulus and the uncertainty in the distance modulus, respectively. Furthermore,    the parameters in the
cosmological models are indicated by  $\theta$.  For example, for the power law reconstruction scheme it given by  $m$,
the exponent of in the $q$-de Sitter it is given by the  non-extensivity parameter $q$. Now we obtain,
\begin{align}
\chi_{\rm SN}^2(\theta)=A(\theta)-\frac{B(\theta)^2}{C(\theta)}\,,
\end{align}
where
\begin{align}
&A(\theta) =\sum_{i=1}^{580}
\frac{\left[\mu_{th}(z_i,\mu_0=0,\theta)-\mu_{obs}(z_i)\right]^2}{
\sigma_\mu(z_i)^2}\,, \\
&B(\theta) =\sum_{i=1}^{580}
\frac{\mu_{th}(z_i,\mu_0=0,\theta)-\mu_{obs}(z_i)}{\sigma_\mu(z_i)^2}\,, \\
&C(\theta) =\sum_{i=1}^{580} \frac{1}{\sigma_\mu(z_i)^2}\,.
\end{align}
If we  use BAO data of $\frac{d_A(z_\star)}{D_V(Z_{BAO})}$,  we have $z_\star \approx 1091$ as
the decoupling time,  $d_A(z)=\int_0^z \frac{dz'}{H(z')}$ as the co-moving angular-diameter
distance and  $D_V(z)=\left(d_A(z)^2\frac{z}{H(z)}\right)^{\frac{1}{3}}$ as the dilation scale.
Using this data set, the  $\chi_\mathrm{BAO}^2$ is defined as
\begin{equation}
 \chi_{\rm BAO}^2=X^T C^{-1} X\,.
\end{equation}
Here what is needed is    in the following column vector,
\begin{equation}
X=\left( \begin{array}{c}
        \frac{d_A(z_\star)}{D_V(0.106)} - 30.95 \\
        \frac{d_A(z_\star)}{D_V(0.2)} - 17.55 \\
        \frac{d_A(z_\star)}{D_V(0.35)} - 10.11 \\
        \frac{d_A(z_\star)}{D_V(0.44)} - 8.44 \\
        \frac{d_A(z_\star)}{D_V(0.6)} - 6.69 \\
        \frac{d_A(z_\star)}{D_V(0.73)} - 5.45
        \end{array} \right)\,,
\end{equation}
Furthermore, the  $C^{-1}$ is  the inverse covariance matrix. Finally,
we use the observational data on Hubble parameter as recently
compiled by \cite{Farooq:2013hq}  in the redshift range $0.07\leq z \leq 2.3 $.  In this data set,
the Hubble constant $H_0$ is taken from the
 PLANCK 2013 results \cite{Planck}. It may be noted that the
 the normalized Hubble parameter  is defined by $h=H/H_0$. In this data set,
the $\chi^2$ for the normalized Hubble parameter is computed as
\begin{align}
\chi_{\rm Hub}^2(\theta)=\sum_{i=1}^{29}
\frac{\left[h_{\rm th}(z_i,\theta)-h_{\rm obs}(z_i)\right]^2}{
\sigma_h(z_i)^2}\,,
\end{align}
where $h_{\rm obs}$ is the observed value of the normalized Hubble parameter, and $h_{\rm th}$ is theoretical values of
the normalized Hubble parameter. The error can now be estimated as
\begin{align}
\sigma_h = \left( \frac{\sigma_H}{H}+\frac{\sigma_{H_0}}{H_0} \right) h,
\label{errorh}
\end{align}
where $\sigma_H$  is the error in $H$, and $\sigma_{H_0}$ is the error in ${H_0}$.

\begin{figure}[hc] \centering
\includegraphics[width=8cm]{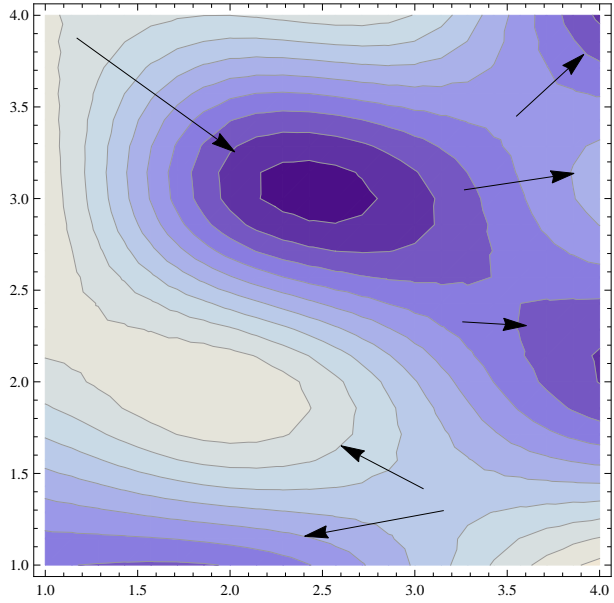}
 \caption{This figure shows the $1\sigma$ (plotted in dark) and  $2\sigma$ (plotted in light) likelihood contours for different cosmological  with joint data
 (SnIa+Hubble+BAO). }\label{likelihood}
\end{figure}
In Figure \ref{likelihood}, we plot the $1\sigma$ (dark regions) and  $2\sigma$ (light regions) likelihood contours for these cosmological models,
Using  the joint data (SNIa+Hubble+BAO), we observe that the best fit value of the parameters which are found to be
$\Omega_{m0} = 0.319$ .
Thus, for models with a power-law the best fit occurs for
 $m = 3.218^{+0.0763}_{-0.0564}(1\sigma)\,^{+0.2134}_{-0.0197}(2\sigma)$. Furthermore, it is possible to have analyse certain models with a future singularity
 after finite time, and for these models, the best fit occurs for  $n = 4.017^{+0.0765}_{-0.0453}(1\sigma)\,^{+0.2341}_{-0.0876}(2\sigma)$.
 The best fit for emergent universe occurs for   $n = 2.054^{+0.0364}_{-0.0312}(1\sigma)\,^{+0.1268}_{-0.0654}(2\sigma),\lambda=6^{+0.0131}_{-0.0976}(1\sigma)\,
 ^{+0.1354}_{-0.0584}(2\sigma)$, and the best fir for
intermediate inflation occurs for  $B = 2.036^{+0.0184}_{-0.0211}(1\sigma)\,^{+0.1287}_{-0.0465}(2\sigma),\theta=0.756^{+0.0123}_{-0.0765}(1\sigma)\,
^{+0.1254}_{-0.0512}(2\sigma)$. Thus, we have analysed different cosmological models in Einstein-aether gravity, and used observational data to analyze the value of
parameters in these cosmological models.

\section{Conclusions}
In this paper, we analysed various different cosmological models based on the
    Einstein-aether gravity.  In Einstein-aether gravity, a time-like vector field couples the usual Einstein Lagrangian, and this time-like vector field
    breaks the Lorentz symmetry of the theory. In this paper, we have analysed various different cosmological models using Einstein-aether gravity. It was demonstrated
    that the aether field modifies the cosmology in a non-trivial way. Explicit expressions for such a modification to various different cosmological models
    were derived  in this paper. Furthermore, the cosmological models based on Einstein-aether gravity were also compared with observational data.
    This was done by using  the cosmographic analysis involving  the {\it Om} parametrization.
Thus,   the SnIa, BAO and Hubble
data was used to obtain    the $1\sigma$ and $2\sigma$ contours for density parameter  $\Omega_m$    arising from the Sne Ia $+$ BAO.

    It is important to perform such an analysis as it is expected that gravitational waves can be used to test Einstein-aether gravity, and as gravitational 
    wave will be used to test several of the predictions of   Einstein-aether gravity, in near future, it is important to analyse the effect of Einstein-aether 
    gravity on cosmology.
      In fact, it has been predicted that gravitational wave detectors can be used to test Einstein-aether gravity \cite{6a}. Thus, it becomes important to 
      analyse various different cosmological models using Einstein-aether gravity. It may be noted that as the Einstein-aether gravity modifies the cosmological 
      models in a non-trivial way, it would also be interesting to analyse quantum cosmology using these modified cosmological models. It would be possible to 
      calculate the Wheeler-DeWitt equation for these cosmological models, and the wave function of the universe can then be obtained as a solution to the 
      Wheeler-DeWitt equation. We would like to mention, that such an analysis would be very interesting and important. Furthermore, as the time-like vector 
      field breaks the time-reparametrization symmetry, it would modify the Wheeler-DeWitt equation in a very non-trivial way. It might be possible to use this 
      time-like aether vector field to obtain  a direction of time, even in the Wheeler-DeWitt equation. Thus, it might be possible that this formalism can be used 
as a solution to the problem of time. It would be interesting to perform such an analysis, for these cosmological models.

It may be noted that the  Horava-Lifshitz gravity has been used for analyzing type IIA string theory \cite{A12}, type  IIB string theory \cite{B12},
AdS/CFT correspondence \cite{ho12, h12, h22,  oh12}, dilaton black branes \cite{d12, d21}, and dilaton black holes \cite{dh12, hd12}.
 As the Horava-Lifshitz gravity is related to the Einstein-aether gravity \cite{2a}, it would be interesting to analyse these systems  using Einstein-aether gravity.
In fact, it has been demonstrated that Einstein-aether gravity can  be related to
  the noncritical string \cite{nonc}. Thus, it would be interesting to analyse this connection further, and also study various cosmological models
  motivated from string theory in Einstein-aether gravity. It may be noted that the
  Einstein-aether gravity has been demonstrated to be equivalent to generalization of Horava-Lifshitz gravity

\section{Appendix}
In this appendix, we will explicitly calculate various cosmological solutions in Einstein-aether gravity.

The first model which we are  studying is the   power law,
\begin{eqnarray}
a\left( t \right) = a_0 t^m. \label{21}
\end{eqnarray}
where $a_0$ is the present day value of $a\left(t\right)$, and we must have $m>0$ for  an accelerating  universe \cite{Rani:2014sia}.
With the choice of scale factor made in Eq. (\ref{21}), we obtain that the Hubble parameter $H$,
\begin{eqnarray}
H = \frac{\dot{a}\left( t \right) }{a\left( t \right) } = \frac{m}{t}. \label{22}
\end{eqnarray}
Moreover, we have that the first and the second time derivatives of the Hubble parameter obtained in Eq. (\ref{22}),
\begin{eqnarray}
\dot{H} &=&  -\frac{m}{t^2}, \label{23} \\
\ddot{H} &=&  \frac{2 m}{t^3} .   \label{24}
\end{eqnarray}
Furthermore, using in Eq. (\ref{13}) the expression of $H$ derived in Eq. (\ref{22}), we obtain that the expression of $K$,
\begin{eqnarray}
K= \frac{3 \varepsilon  m^2}{M^2 t^2}.     \label{25}
\end{eqnarray}
Using in the general expression of $L_{GO}$ given in Eq. (\ref{deflgo}) the expressions of $H$ and $\dot{H}$ obtained in Eqs. (\ref{22}) and (\ref{23}),
we have,
\begin{eqnarray}
L_{GO} =  \frac{t}{\sqrt{m (m \alpha -\beta )}}. \label{26}
\end{eqnarray}
Therefore, we can conclude that the expression of $\rho_{EA}$ with Granda-Oliveros cut-off can be written as
\begin{eqnarray}
\rho_{EA} = \frac{3 c^2 m (m \alpha -\beta )}{t^2}. \label{27}
\end{eqnarray}
Using in Eq. (\ref{18a}) the expressions of $H$ and $\rho_{EA}$ given in Eqs. (\ref{22}) and (\ref{27}) or equivalently in Eq. (\ref{18b}),
the expressions of $K$ and $\rho_{EA}$ given in Eqs. (\ref{25}) and (\ref{27}), we obtain the following differential equation for $F\left( K\right)$,
\begin{eqnarray}
\frac{dF\left( K\right)}{dK} - \frac{F\left( K\right)}{2K} -\frac{c^2 (m \alpha -\beta )}{ \varepsilon  m} =0, \label{28}
\end{eqnarray}
which has the following solution,
\begin{eqnarray}
F\left(K\right) = \frac{2 c^2 K (m \alpha -\beta )}{ \varepsilon  m}+C_1\sqrt{K} , \label{29}
\end{eqnarray}
where $C_1$ represents an integration constant.\\
Using in Eq. (\ref{19}) the expression of $\rho$ derived in Eq. (\ref{27}) along with the expression of $H$ obtained in Eq. (\ref{22}),
we can write the pressure $p_{EA}$ as
\begin{eqnarray}
 p_{EA} = \frac{c^2\left(2-3m \right)\left( m\alpha - \beta  \right)}{t^2}. \label{30}
\end{eqnarray}
Therefore, the EoS parameter $\omega_{EA}$ is given by
\begin{eqnarray}
\omega_{EA} = \frac{p_{EA}}{\rho_{EA}} =-1 + \frac{2}{3m}. \label{31}
\end{eqnarray}
At Ricci scale,   for $\alpha =2$ and $\beta =1$, we obtain,
\begin{eqnarray}
L_{GO} &=&  \frac{t}{\sqrt{m (2m  -1 )}}, \label{26a}\\
\rho_{EA} &=& \frac{3 c^2 m (2m  -1 )}{t^2}, \label{27a}\\
F\left(K\right) &=& \frac{2 c^2 K (2m  -1 )}{ \varepsilon  m}+C_1\sqrt{K} , \label{29a}\\
 p_{EA} &=& \frac{c^2\left(2-3m \right)\left( 2m - 1  \right)}{t^2}. \label{30a}
\end{eqnarray}
Moreover, for $\alpha 	\approx 0.8824$ and $\beta	\approx 0.5016$, i.e., for the values of $\alpha$ and $\beta$ corresponding to a non-flat Universe, we obtain,
\begin{eqnarray}
L_{GO} &=&  \frac{t}{\sqrt{m ( 0.8824m  -0.5016 )}}, \label{26b}\\
\rho_{EA} &=& \frac{3 c^2 m ( 0.8824m  -0.5016 )}{t^2}, \label{27b}\\
F\left(K\right) &=& \frac{2 c^2 K ( 0.8824m  -0.5016 )}{ \varepsilon  m}+C_1\sqrt{K} , \label{29b}\\
 p_{EA} &=& \frac{c^2\left(2-3m \right)\left( 0.8824m - 0.5016  \right)}{t^2}. \label{30b}
\end{eqnarray}
Furthermore,   for   $\alpha 	\approx 0.8502$ and $\beta	\approx 0.4817$, i.e.,
for the values of $\alpha$ and $\beta$ corresponding to a  flat Universe, we obtain,
\begin{eqnarray}
L_{GO} &=&  \frac{t}{\sqrt{m ( 0.8502m  -0.4817 )}}, \label{26c}\\
\rho_{EA} &=& \frac{3 c^2 m ( 0.8502m  -0.4817 )}{t^2}, \label{27c}\\
F\left(K\right) &=& \frac{2 c^2 K ( 0.8502m  -0.4817 )}{ \varepsilon  m}+C_1\sqrt{K} , \label{29c}\\
 p_{EA} &=& \frac{c^2\left(2-3m \right)\left( 0.8502m - 0.4817  \right)}{t^2}. \label{30c}
\end{eqnarray}

We now consider the Chen-Jing model studied in this paper, i.e., the one with the first and the second time derivatives of the Hubble parameter $H$.
Using  the expressions of $H$, $\dot{H}$ and $\ddot{H}$ given in Eqs. (\ref{22}), (\ref{23}) and (\ref{24}) in Eq. (\ref{rho}), we obtain the expression of
$\rho_{EA}$ with higher derivatives of the Hubble parameter,
\begin{eqnarray}
\rho_{EA} = \frac{3 c^2\left[2 \alpha +m (m \beta -\gamma )\right]}{t^2} . \label{32}
\end{eqnarray}
Using the expressions of $H$ and $\rho_{EA}$ given in Eqs. (\ref{22}) and (\ref{32}) in Eq. (\ref{18a}) or equivalently in Eq. (\ref{18b}),
the expressions of $K$ and $\rho_{EA}$ from Eqs. (\ref{25}) and (\ref{32}), we obtain the following differential equation for $F\left( K\right)$,
\begin{eqnarray}
\frac{dF\left( K\right)}{dK} - \frac{F\left( K\right)}{2K} -\frac{c^2 \left[2 \alpha +m (m \beta -\gamma )\right]}{ \varepsilon  m^2} =0,   \label{33}
\end{eqnarray}
whose solution is given by
\begin{eqnarray}
F\left(K\right) = \frac{2 c^2 K \left[2 \alpha +m (m \beta -\gamma )\right]}{ \varepsilon  m^2}+C_2\sqrt{K},  \label{34}
\end{eqnarray}
where $C_2$ is an integration constant.
Substituting in Eq. (\ref{19}) the expression of $\rho_{EA}$ derived in Eq. (\ref{32}),
along with the expression of $H$ obtained in Eq. (\ref{22}), we can write the pressure $p_{EA}$ as
\begin{eqnarray}
p_{EA} = -\frac{c^2 (-2+3 m) \left[2 \alpha +m \left(m \beta -\gamma \right)\right]}{m t^2}. \label{35}
\end{eqnarray}
Therefore, the EoS parameter $\omega_{EA}$ is given by
\begin{eqnarray}
\omega_{EA} = \frac{p_{EA}}{\rho_{EA}} =- 1 + \frac{2}{3m},  \label{36}
\end{eqnarray}
which is the same  as   Eq. (\ref{31}).

We now consider the second scale factor considered in this work, which is another form of the    power law \cite{a, za},
\begin{eqnarray}
a\left( t \right) = a_0 (t_s  -t )^{-n} \label{37},
\end{eqnarray}
where $n>0$ and $t<t_s$. Here $a_0$ is  the present day value of $a(t)$,  while $t_s$ is the probable future singularity at finite time.
So, this model has a future singularity.
With the choice of scale factor made in Eq. (\ref{37}), we obtain   the Hubble parameter $H$,
\begin{eqnarray}
H = \frac{\dot{a}\left( t \right) }{a\left( t \right) } = \frac{n}{t_s -t}.  \label{38}
\end{eqnarray}
Moreover,  the first and the second time derivatives of the Hubble parameter are given by
\begin{eqnarray}
\dot{H} &=& \frac{n}{(t_s - t )^2},    \label{39}\\
\ddot{H}  &=& \frac{2 n}{(t_s - t )^3}.  \label{40}
\end{eqnarray}
Furthermore, using in  Eq. (\ref{13})   the expression of $H$ derived in Eq. (\ref{38}), we obtain that the expression of $K$,
\begin{eqnarray}
K= \frac{3 \varepsilon  n^2}{M^2 (t_s - t )^2}.  \label{41}
\end{eqnarray}
Using in the general expression of $L_{GO}$ given in Eq. (\ref{deflgo}),  the expressions of $H$ and $\dot{H}$ obtained in Eqs. (\ref{38}) and (\ref{39}),
we obtain
\begin{eqnarray}
L_{GO}  = \frac{\left( t_s -t \right)}{\sqrt{n (n \alpha +\beta )}}. \label{42}
\end{eqnarray}
Therefore, we can conclude that the expression of $\rho_{EA}$ with Granda-Oliveros cut-off can be written as
\begin{eqnarray}
\rho_{EA} =  \frac{3 c^2 n (n \alpha +\beta )}{(t-t_s )^2}. \label{43}
\end{eqnarray}
Using in Eq. (\ref{18a}) the expressions of $H$ and $\rho_{EA}$ given in Eqs. (\ref{38}) and (\ref{43}),
or equivalently in Eq. (\ref{18b}) the expressions of $K$ and $\rho_{EA}$ given in Eqs. (\ref{41}) and (\ref{43}),
we obtain the following differential equation for $F\left( K\right)$,
\begin{eqnarray}
\frac{dF}{dK} - \frac{F}{2K} -\frac{c^2 (n \alpha +\beta )}{ \varepsilon  n}=0, \label{44}
\end{eqnarray}
whose solution is given by:
\begin{eqnarray}
F\left( K\right) =  \frac{2 c^2 K (n \alpha +\beta )}{ \varepsilon  n}+C_3\sqrt{K}, \label{45}
\end{eqnarray}
where $C_3$ represents an integration constant.

Using in Eq. (\ref{19}) the expression of $\rho_{EA}$ derived in Eq. (\ref{43}),  along with the expression of $H$ obtained in Eq. (\ref{43}),
we can write the pressure $p_{EA}$ as
\begin{eqnarray}
p_{EA}=\frac{c^2\left(2+3n \right)\left( n\alpha + \beta  \right)}{\left(t_s -t\right)^2}. \label{46}
\end{eqnarray}
Therefore, the EoS parameter $\omega_{EA}$ is given by
\begin{eqnarray}
\omega_{EA} = \frac{p_{EA}}{\rho_{EA}}=- 1 - \frac{2}{3n}. \label{47}
\end{eqnarray}
At  Ricci scale, i.e.,  for $\alpha =2$ and $\beta =1$, we obtain
\begin{eqnarray}
L_{GO}  &=& \frac{\left( t_s -t \right)}{\sqrt{n (2n  +1 )}}, \label{42a}\\
\rho_{EA} &=&  \frac{3 c^2 n (2n  +1 )}{(t-t_s )^2}, \label{43a}\\
F\left( K\right) &=&  \frac{2 c^2 K (2n  +1 )}{ \varepsilon  n}+C_3\sqrt{K}, \label{45a}\\
p_{EA}&=&\frac{c^2\left(2+3n \right)\left( 2n + 1  \right)}{\left(t_s -t\right)^2}. \label{46a}
\end{eqnarray}
Moreover, for $\alpha \approx 0.8824$ and $\beta\approx 0.5016$, i.e., for the value of $\alpha$ and $\beta$ corresponding to a non-flat universe, we obtain
\begin{eqnarray}
L_{GO}  &=& \frac{\left( t_s -t \right)}{\sqrt{n (0.8824n +0.5016 )}}, \label{42b}\\
\rho_{EA} &=&  \frac{3 c^2 n (0.8824n  +0.5016)}{(t-t_s )^2}, \label{43b}\\
F\left( K\right) &=&  \frac{2 c^2 K (0.8824n +0.5016 )}{ \varepsilon  n}+C_3\sqrt{K},  \label{45b}\\
p_{EA}&=&\frac{c^2\left(2+3n \right)\left( 0.8824n+ 0.5016  \right)}{\left(t_s -t\right)^2}. \label{46b}
\end{eqnarray}
Furthermore, for $\alpha \approx 0.8502$ and $\beta\approx 0.4817$, i.e., for the value of $\alpha$ and $\beta$ corresponding to a  flat Universe, we obtain
\begin{eqnarray}
L_{GO}  &=& \frac{\left( t_s -t \right)}{\sqrt{n (0.8502n +0.4817 )}}, \label{42c}\\
\rho_{EA} &=&  \frac{3 c^2 n (0.8502n  +0.4817)}{(t-t_s )^2}, \label{43c}\\
F\left( K\right) &=&  \frac{2 c^2 K (0.8502n +0.4817 )}{ \varepsilon  n}+C_3\sqrt{K},  \label{45c}\\
p_{EA}&=&\frac{c^2\left(2+3n \right)\left( 0.8502n+ 0.4817  \right)}{\left(t_s -t\right)^2}. \label{46c}
\end{eqnarray}

We now consider the Chen-Jing model studied in this paper, i.e., the energy density model with the first and the second time derivatives of the Hubble parameter $H$.
Using in Eq. (\ref{rho}) the expressions of $H$, $\dot{H}$ and $\ddot{H}$ obtained in Eqs. (\ref{38}), (\ref{39}) and (\ref{40}),
we obtain   the expression for  $\rho_{EA}$,
\begin{eqnarray}
\rho_{EA} = \frac{3 c^2}{(t_s - t )^2} \left[ 2 \alpha +n (n \beta +\gamma )  \right]. \label{48}
\end{eqnarray}
Using in Eq. (\ref{18a}) the expressions of $H$ and $\rho_{EA}$ given in Eqs. (\ref{38}) and (\ref{48}),
or equivalently in Eq. (\ref{18b}) the expressions of $K$ and $\rho_{EA}$ given in Eqs. (\ref{41}) and (\ref{48}),
we obtain the following differential equation for $F\left( K\right)$,
\begin{eqnarray}
 \frac{dF}{dK} - \frac{F}{2K}-\frac{c^2 \left[2 \alpha +n (n \beta +\gamma )\right]}{ \varepsilon  n^2}=0, \label{49}
\end{eqnarray}
which solution is given by,
\begin{eqnarray}
F\left( K\right) =\frac{2 c^2 K \left[2 \alpha +n (n \beta +\gamma )\right]}{ \varepsilon  n^2}+C_4\sqrt{K},  \label{50}
\end{eqnarray}
where $C_4$ represents an integration constant.

Using in Eq. (\ref{19}) the expression of $\rho_{EA}$ derived in Eq. (\ref{48}) along with the expression of $H$ obtained in Eq. (\ref{38}),
we can write the pressure $p_{EA}$ as follows,
\begin{eqnarray}
p_{EA}= -\frac{c^2 (2+3 n) \left[2 \alpha +n (n \beta +\gamma )\right]}{n \left(t-t_s \right)^2}. \label{51}
\end{eqnarray}
Therefore, the EoS parameter $\omega_{EA}$ is given by,
\begin{eqnarray}
\omega_{EA}= \frac{p_{EA}}{\rho_{EA}} =  - 1 - \frac{2}{3n}, \label{52}
\end{eqnarray}
which is the same results of Eq. (\ref{47}).

 We can also analyse an
  emergent universe  using this analysis. The scale factor for such  a cosmological model is given by \cite{17, 30},
\begin{eqnarray}
a\left( t \right) =  A \left(B+e^{n t}\right)^{\lambda } \label{53}
\end{eqnarray}
where $A$, $B$, $n$ and $\lambda$ are four positive constant parameters.
With the choice of scale factor given in Eq. (\ref{53}), we can   obtain   the Hubble parameter $H$,
\begin{eqnarray}
H =  \frac{e^{n t} n \lambda }{B+e^{n t}}. \label{54}
\end{eqnarray}
Moreover, the first and the second time derivatives of the Hubble parameter are given by
\begin{eqnarray}
\dot{H} &=& \frac{B e^{n t} n^2 \lambda }{\left(B+e^{n t}\right)^2},   \label{55}\\
\ddot{H} &=& \frac{B e^{n t} \left(B-e^{n t}\right) n^3 \lambda }{\left(B+e^{n t}\right)^3}. \label{56}
\end{eqnarray}
Furthermore, using in Eq. (\ref{13}) the expression of $H$ derived in Eq. (\ref{54}), we obtain that the expression of $K$
\begin{eqnarray}
K=  \frac{3 \varepsilon  e^{2 n t} n^2 \lambda ^2}{\left(B+e^{n t}\right)^2 M^2}. \label{57}
\end{eqnarray}
Using in the general expression of $L_{GO}$ given in Eq. (\ref{deflgo}) the expressions of $H$ and $\dot{H}$ obtained in Eqs. (\ref{54}) and (\ref{55}),
we obtain
\begin{eqnarray}
L_{GO} = \frac{\left(B+e^{n t}\right)}{\sqrt{e^{n t} n^2 \lambda  \left(B \beta +e^{n t} \alpha  \lambda \right)}}. \label{58}
\end{eqnarray}
Therefore, we can conclude that the expression of $\rho_{EA}$ with Granda-Oliveros cut-off can be written as
\begin{eqnarray}
\rho_{EA} = \frac{3 c^2 e^{n t} n^2 \lambda  \left(B \beta +e^{n t} \alpha  \lambda \right)}{\left(B+e^{n t}\right)^2}. \label{59}
\end{eqnarray}
Using in Eq. (\ref{18a}) the expressions of $H$ and $\rho_{EA}$ given in Eqs. (\ref{54}) and (\ref{59}),
or equivalently in Eq. (\ref{18b}) the expressions of $K$ and $\rho_{EA}$ given in Eqs. (\ref{57}) and (\ref{59}),
we obtain the following differential equation for $F\left( K\right)$,
\begin{eqnarray}
\frac{dF}{dK} - \frac{F}{2K} -\frac{c^2 \left\{\alpha  \lambda +\beta  \left[-1+\left(\frac{3 \varepsilon   n^2 \lambda ^2}{KM^2}\right)^{1/2}\right]\right\}}{ \varepsilon  \lambda }=0, \label{60}
\end{eqnarray}
which has a solution given by
\begin{eqnarray}
F\left(K\right) =\sqrt{K} C_5+\frac{c^2 K \left(2 \alpha \lambda -2 \beta + \beta  \sqrt{\frac{3\varepsilon n^2
\lambda ^2}{K M^2}} \log K\right)}{\varepsilon  \lambda} , \label{61}
\end{eqnarray}
where $C_5$ represents an integration constant.

Using in Eq. (\ref{19}) the expression of $\rho_{EA}$ derived in Eq. (\ref{59}) along with the expression of $H$ obtained in Eq. (\ref{54}),
we can write the pressure $p_{EA}$ as
\begin{eqnarray}
p_{EA} =-\frac{c^2 n^2 \left\{B^2 \beta +3 e^{2 n t} \alpha  \lambda ^2+B e^{n t} \left[2 \alpha  \lambda +\beta  \left(-1+3 \lambda \right)\right]\right\}}{\left(B+e^{nt}\right)^2}. \label{62}
\end{eqnarray}
Therefore, the EoS parameter $\omega_{EA}$ is given by
\begin{eqnarray}
\omega_{EA} = \frac{p_{EA}}{\rho_{EA}} =-1 +\frac{B \left(\beta -B e^{-n t} \beta -2 \alpha  \lambda \right)}{3\lambda  \left(B \beta +e^{n t} \alpha  \lambda\right)}. \label{63}
\end{eqnarray}
At Ricci scale, i.e., for $\alpha =2$ and $\beta =1$, we obtain
\begin{eqnarray}
L_{GO} &=& \frac{\left(B+e^{n t}\right)}{\sqrt{e^{n t} n^2 \lambda  \left(B  +2e^{n t}   \lambda \right)}}, \label{58a}\\
\rho_{EA} &=& \frac{3 c^2 e^{n t} n^2 \lambda  \left(B  +2e^{n t}   \lambda \right)}{\left(B+e^{n t}\right)^2}, \label{59a}\\
F\left(K\right) &=&\sqrt{K} C_5+\frac{c^2 K \left(4 \lambda -2  +   \sqrt{\frac{3\varepsilon n^2\lambda ^2}{K M^2}} \log K\right)}{\varepsilon  \lambda}, \label{61a}\\
p_{EA} &=&-\frac{c^2 n^2 \left[B^2  +6 e^{2 n t}  \lambda ^2+B e^{n t} (7 \lambda -1)\right]}{\left(B+e^{nt}\right)^2}, \label{62a}\\
\omega_{EA} &=& -1 +\frac{B \left(1 -B e^{-n t}  -4  \lambda \right)}{3\lambda  \left(B +2e^{n t}   \lambda\right)}. \label{63a}
\end{eqnarray}
For $\alpha \approx 0.8824$ and $\beta \approx 0.5016$, i.e., for the value of $\alpha$ and $\beta$ corresponding to a non-flat Universe, we obtain
\begin{eqnarray}
L_{GO} &=& \frac{\left(B+e^{n t}\right)}{\sqrt{e^{n t} n^2 \lambda  \left(0.5016B  +0.8824e^{n t}   \lambda \right)}}, \label{58b}\\
\rho_{EA} &=& \frac{3 c^2 e^{n t} n^2 \lambda  \left(0.5016B  +0.8824e^{n t}  \lambda \right)}{\left(B+e^{n t}\right)^2}, \label{59b}\\
F\left(K\right) &=&\sqrt{K} C_5+\frac{c^2 K \left(1.7648 \lambda -1.0032 + 0.5016  \sqrt{\frac{3\varepsilon n^2 \lambda ^2}{K M^2}} \log K\right)}{\varepsilon  \lambda},  \label{61b}\\
p_{EA} &=&-\frac{c^2 n^2 \left\{0.5016B^2  +2.6472 e^{2 n t}  \lambda ^2+B e^{n t} \left[1.7648 \lambda +0.5016  \left(-1+3 \lambda \right)\right]\right\}}{\left(B+e^{nt}\right)^2} \label{62b},\\
\omega_{EA} &=& -1 +\frac{B \left(0.5016 -0.5016B e^{-n t}  -1.7648  \lambda \right)}{3\lambda  \left(0.5016B  +0.8824e^{n t}   \lambda\right)}. \label{63b}
\end{eqnarray}
For $\alpha \approx 0.8502$ and $\beta \approx 0.4817$, i.e., for the value of $\alpha$ and $\beta$ corresponding to a  flat Universe, we obtain
\begin{eqnarray}
L_{GO} &=& \frac{\left(B+e^{n t}\right)}{\sqrt{e^{n t} n^2 \lambda  \left(0.4817B  +0.8502e^{n t}   \lambda \right)}}, \label{58b}\\
\rho_{EA} &=& \frac{3 c^2 e^{n t} n^2 \lambda  \left(0.4817B  +0.8502e^{n t}  \lambda \right)}{\left(B+e^{n t}\right)^2}, \label{59b}\\
F\left(K\right) &=&\sqrt{K} C_5+\frac{c^2 K \left(1.7004 \lambda -0.9634 + 0.4817  \sqrt{\frac{3\varepsilon n^2 \lambda ^2}{K M^2}} \log K\right)}{\varepsilon  \lambda},  \label{61b}\\
p_{EA} &=&-\frac{c^2 n^2 \left\{0.4817B^2  +2.5506 e^{2 n t}  \lambda ^2+B e^{n t} \left[1.7004 \lambda +0.4817  \left(-1+3 \lambda \right)\right]\right\}}{\left(B+e^{nt}\right)^2} \label{62b},\\
\omega_{EA} &=& -1 +\frac{B \left(0.4817 -0.4817B e^{-n t}  -1.7004  \lambda \right)}{3\lambda  \left(0.4817B  +0.8502e^{n t}   \lambda\right)}. \label{63b}
\end{eqnarray}

We now consider the Chen-Jing model studied in this paper, i.e., the energy density model with the first and the second time derivatives of the Hubble parameter $H$.
Using in Eq. (\ref{rho}) the expressions of $H$, $\dot{H}$ and $\ddot{H}$ obtained in Eqs. (\ref{54}), (\ref{55}) and (\ref{56}),
we obtain that the expression of $\rho_{EA}$ is given by
\begin{eqnarray}
\rho_{EA} = \frac{3 c^2}{\left(B+e^{n t}\right)^2} \left[B \left(B-e^{n t}\right) n^2 \alpha +  B e^{n t} n^2 \gamma \lambda +e^{2n t} n^2 \beta \lambda ^2\right]. \label{64}
\end{eqnarray}
Using in Eq. (\ref{18a}) the expressions of $H$ and $\rho_{EA}$ given in Eqs. (\ref{54}) and (\ref{64}),
or equivalently in Eq. (\ref{18b}) the expressions of $K$ and $\rho_{EA}$ given in Eqs. (\ref{57}) and (\ref{64}),
we obtain the following differential equation for $F\left( K\right)$,
\begin{eqnarray}
\frac{dF}{dK} &-& \frac{F}{2K} - \frac{c^2}{3 \varepsilon  \lambda ^2}\times \nonumber \\
&& \left\{\beta  \lambda ^2-\left(\alpha -\gamma  \lambda \right) \left[-1+\left(\frac{3 \varepsilon   n^2 \lambda ^2}{KM^2}\right)^{1/2}\right] \right. \nonumber \\
&&\left. +\alpha  \left[-1+\left(\frac{3
\varepsilon   n^2 \lambda ^2}{KM^2}\right)^{1/2}\right]^2\right\}=0, \label{65}
\end{eqnarray}
which solution is given by
\begin{eqnarray}
F\left(K\right) &=& \frac{A_1}{B_1} + C_6\sqrt{K}, \label{66}
\end{eqnarray}
where $C_6$ represents an integration constant,  $A_1$ and $B_1$ are given by
\begin{eqnarray}
A_1 &=& c^2 \left[-6 \varepsilon n^2 \alpha  \lambda ^2+2 K M^2 \left(2 \alpha -\gamma  \lambda+\beta  \lambda ^2\right)  \right. \nonumber \\
&&\left. +\sqrt{3} K M^2 \sqrt{\frac{\varepsilon n^2 \lambda ^2}{K M^2}} \left(-3 \alpha +\gamma  \lambda \right) \log K\right],\\
B_1 &=& \varepsilon M^2 \lambda ^2.
\end{eqnarray}
Using in Eq. (\ref{19}) the expression of $\rho_{EA}$ derived in Eq. (\ref{64})
along with the expression of $H$ obtained in Eq. (\ref{54}), we can write the pressure $p_{EA}$ as
\begin{eqnarray}
p_{EA} = \frac{A_2}{B_2}, \label{67}
\end{eqnarray}
where
\begin{eqnarray}
A_2 &=& c^2 n^2 \left\{-3 e^{2 n t} \beta  \lambda ^3-B^2 (3 \alpha  (-1+\lambda )+\gamma  \lambda ) \right. \nonumber \\
&&\left.+B e^{n t} \left[\alpha  (-1+3 \lambda )+\lambda(\gamma -2 \beta  \lambda -3 \gamma  \lambda )\right]\right\},\\
B_2 &=&  \left(B+e^{n t}\right)^2 \lambda.
\end{eqnarray}
Therefore, the EoS parameter $\omega_{EA}$ is given by
\begin{eqnarray}
\omega_{EA} = \frac{p_{EA}}{\rho_{EA}} =-1+\frac{B \left\{B (3 \alpha -\gamma  \lambda )-e^{n t} \left[\alpha +\lambda  \left(-\gamma +2 \beta  \lambda \right)\right]\right\}}{3 \lambda  \left[B^2 \alpha
+e^{2 n t} \beta  \lambda ^2+B e^{n t} (-\alpha +\gamma  \lambda )\right]}. \label{68}
\end{eqnarray}

We now consider the   scale factor in the  intermediate inflation  \cite{18, 31}:
\begin{eqnarray}
a\left( t \right) =  e^{B t^{\theta }}, \label{69}
\end{eqnarray}
where $B >0$ and $0< \theta <1$.
With the choice of scale factor given in Eq. (\ref{69}), we obtain that the Hubble parameter $H$,
\begin{eqnarray}
H  = B\theta t^{-1+\theta }. \label{70}
\end{eqnarray}
Moreover, we have that the first and the second time derivatives of the Hubble parameter are given by
\begin{eqnarray}
\dot{H} &=&  B  (-1+\theta ) \theta t^{-2+\theta },  \label{71}\\
\ddot{H} &=&  B  (-2+\theta ) (-1+\theta ) \theta t^{-3+\theta }. \label{72}
\end{eqnarray}
Furthermore, using in Eq. (\ref{13}) the expression of $H$ derived in Eq. (\ref{70}), we obtain that the expression of $K$
\begin{eqnarray}
K= \frac{3 B^2 \varepsilon  t^{-2+2 \theta } \theta ^2}{M^2}. \label{73}
\end{eqnarray}
Using in the general expression of $L_{GO}$ given in Eq. (\ref{deflgo}),
the expressions of $H$ and $\dot{H}$ obtained in Eqs. (\ref{70}) and (\ref{71}), we obtain
\begin{eqnarray}
L_{GO} = \frac{1}{\sqrt{B t^{-2+\theta } \theta  \left[\beta  (-1+\theta )+B t^{\theta } \alpha  \theta \right]}}. \label{74}
\end{eqnarray}
Therefore, we can conclude that the expression of $\rho_{EA}$ with Granda-Oliveros cut-off can be written as
\begin{eqnarray}
\rho_{EA} = 3 B c^2 t^{-2+\theta } \theta  \left[\beta  (-1+\theta )+B t^{\theta } \alpha  \theta \right]. \label{75}
\end{eqnarray}
Using in Eq. (\ref{18a}) the expressions of $H$ and $\rho_{EA}$ given in Eqs. (\ref{69}) and (\ref{75}),
or equivalently in Eq. (\ref{18b}) the expressions of $K$ and $\rho_{EA}$ given in Eqs. (\ref{73}) and (\ref{75}),
we obtain the following differential equation for $F\left( K\right)$
\begin{eqnarray}
\frac{dF}{dK} - \frac{F}{2K} -\frac{c^2 \left[\beta  \left(-1+\theta \right)  \left(\frac{K M^2}{3B^2 \varepsilon  \theta ^2}\right)^{\frac{-\theta}{2(-1+\theta )}}+B \alpha  \theta \right]}{ B \varepsilon  \theta }=0, \label{76}
\end{eqnarray}
which solution is given by
\begin{eqnarray}
F\left( K \right) = \frac{2 c^2 K   \left(\frac{K M^2}{3B^2 \varepsilon  \theta ^2}\right)^{\frac{-\theta}{2(-1+\theta )}} \left[-\beta  (-1+\theta )^2+B \alpha \theta  \left(\frac{K M^2}{3B^2 \varepsilon  \theta^2}\right)^{\frac{\theta}{2 (-1+\theta )}}\right]  }{ B \varepsilon  \theta }+C_7\sqrt{K}, \label{77}
\end{eqnarray}
where $C_7$ represents an integration constant.
Using in Eq. (\ref{19}) the expression of $\rho_{EA}$ derived in Eq. (\ref{75}) along with the expression of $H$ obtained in Eq. (\ref{70}),
we can write the pressure $p_{EA}$ as
\begin{eqnarray}
p_{EA} = -\frac{c^2 \left\{\beta  (-1+\theta ) \left(-2+\theta +3 B t^{\theta } \theta \right)+B t^{\theta } \alpha  \theta  \left[-2+\left(2+3B t^{\theta }\right) \theta \right]\right\}}{t^2}. \label{78}
\end{eqnarray}
Therefore, the EoS parameter $\omega_{EA}$ is given by
\begin{eqnarray}
\omega_{EA} =  \frac{p_{EA}}{\rho_{EA}} =-1-\frac{t^{-\theta } (-2+\theta )}{3 B \theta }+\frac{\alpha  \theta }{3 \beta -3 B t^{\theta } \alpha  \theta -3 \beta  \theta } \label{79}.
\end{eqnarray}
At Ricci scale, i.e., for $\alpha =2$ and $\beta =1$, we obtain
\begin{eqnarray}
L_{GO} &=& \frac{1}{\sqrt{B t^{-2+\theta } \theta  \left[ (-1+\theta )+2B t^{\theta }   \theta \right]}}, \label{74a}\\
\rho_{EA} &=& 3 B c^2 t^{-2+\theta } \theta  \left[ (-1+\theta )+2B t^{\theta }   \theta \right] \label{75a},\\
F\left( K \right) &=& \frac{2 c^2 K    \left(\frac{K M^2}{3B^2 \varepsilon  \theta ^2}\right)^{\frac{-\theta}{2(-1+\theta )}}    \left[- (-1+\theta )^2+2B\theta \left(\frac{K M^2}{3B^2 \varepsilon  \theta^2}\right)^{\frac{\theta}{2 (-1+\theta )}}  \right]}{ B \varepsilon  \theta }+\sqrt{K}C_7 \label{77a},\\
p_{EA} &=& -\frac{c^2 \left\{ (-1+\theta ) \left(-2+\theta +3 B t^{\theta } \theta \right)+2B t^{\theta } \theta  \left[-2+\left(2+3B t^{\theta }\right) \theta \right]\right\}}{t^2} \label{78a},\\
\omega_{EA} &=& -1-\frac{t^{-\theta } (-2+\theta )}{3 B \theta }+\frac{2 \theta }{3  -6 B t^{\theta }   \theta -3  \theta } \label{79a}.
\end{eqnarray}
For $\alpha \approx 0.8824$ and $\beta \approx 0.5016$, i.e., for the value of $\alpha$ and $\beta$ corresponding to a non-flat Universe, we obtain
\begin{eqnarray}
L_{GO} &=& \frac{1}{\sqrt{B t^{-2+\theta } \theta  \left[0.5016  (-1+\theta )+0.8824B t^{\theta }   \theta \right]}}, \label{74b}\\
\rho_{EA} &=& 3 B c^2 t^{-2+\theta } \theta  \left[0.5016 (-1+\theta )+0.8824B t^{\theta }   \theta \right] \label{75b},\\
F\left( K \right) &=& \frac{2 c^2 K    \left(\frac{K M^2}{3B^2 \varepsilon  \theta ^2}\right)^{\frac{-\theta}{2(-1+\theta )}} \left[-0.5016 (-1+\theta )^2+0.8824B \theta   \left(\frac{K M^2}{3B^2 \varepsilon  \theta^2}\right)^{\frac{\theta}{2 (-1+\theta )}}  \right]}{ B \varepsilon  \theta }\nonumber \\
&&+\sqrt{K}C_7\label{77b},\\
p_{EA} &=& -\frac{c^2}{t^2}\times  \left\{0.5016 (\theta -1) \left(\theta -2+3 B t^{\theta } \theta \right)\right. \nonumber \\
&&\left. +0.8824B t^{\theta }  \theta  \left[\left(2+3B t^{\theta }\right) \theta -2\right]\right\} \label{78b},\\
\omega_{EA} &=& -1-\frac{t^{-\theta } (-2+\theta )}{3 B \theta }+\frac{0.8824  \theta }{1.5048-  2.6472 B t^{\theta }  \theta -1.5048  \theta }. \label{79b}
\end{eqnarray}
Furthermore, for $\alpha \approx 0.8502$ and $\beta\approx 0.4817$, i.e., for the value of $\alpha$ and $\beta$ corresponding to a  flat Universe, we obtain
\begin{eqnarray}
L_{GO} &=& \frac{1}{\sqrt{B t^{-2+\theta } \theta  \left[0.4817  (-1+\theta )+0.8502 t^{\theta }   \theta \right]}}, \label{74c}\\
\rho_{EA} &=& 3 B c^2 t^{-2+\theta } \theta  \left[0.4817 (-1+\theta )+0.8502 t^{\theta }   \theta \right] \label{75c},\\
F\left( K \right) &=& \frac{2 c^2 K    \left(\frac{K M^2}{3B^2 \varepsilon  \theta ^2}\right)^{\frac{-\theta}{2(-1+\theta )}} \left[-0.4817 (-1+\theta )^2+0.8502 \theta   \left(\frac{K M^2}{3B^2 \varepsilon  \theta^2}\right)^{\frac{\theta}{2 (-1+\theta )}}  \right]}{ B \varepsilon  \theta }\nonumber \\
&&+\sqrt{K}C_7\label{77c},\\
p_{EA} &=& -\frac{c^2 \left\{0.4817 (\theta -1) \left(\theta -2+3 B t^{\theta } \theta \right)+0.8502 t^{\theta }  \theta  \left[\left(2+3B t^{\theta }\right) \theta -2\right]\right\}}{t^2} \label{78c},\\
\omega_{EA} &=& -1-\frac{t^{-\theta } (-2+\theta )}{3 B \theta }+\frac{0.8502  \theta }{1.5048-  2.6472 B t^{\theta }  \theta -1.5048  \theta }. \label{79c}
\end{eqnarray}

We now consider the Chen-Jing model studied in this paper, i.e.,  the one with the first and the second time derivatives of the Hubble parameter $H$.
Using in Eq. (\ref{rho}) the expressions of $H$, $\dot{H}$ and $\ddot{H}$ obtained in Eqs. (\ref{70}), (\ref{71}) and (\ref{72}),
we obtain that the expression of $\rho_{EA}$,
\begin{eqnarray}
\rho_{EA} = 3 c^2 \left[\frac{\alpha  (-2+\theta ) (-1+\theta )}{t^2}+B t^{-2+\theta } \gamma  (-1+\theta ) \theta +B^2 t^{-2+2 \theta } \beta  \theta^2\right].\label{80}
\end{eqnarray}
Using in Eq. (\ref{18a}) the expressions of $H$ and $\rho_{EA}$ given in Eqs. (\ref{69}) and (\ref{80}),
or equivalently in Eq. (\ref{18b}) the expressions of $K$ and $\rho_{EA}$ given in Eqs. (\ref{73}) and (\ref{80}),
we obtain the following differential equation for $F\left( K\right)$,
\begin{eqnarray}
\frac{dF}{dK} - \frac{F}{2K} -\frac{A_3}{B_3}=0, \label{81}
\end{eqnarray}
where
\begin{eqnarray}
A_3 &=& 3^{-2^{2 \theta } \left(\frac{1}{-1+\theta }\right)^{-2 \theta }} c^2 \left(\frac{K M^2}{B^2 \varepsilon  \theta ^2}\right)^{2^{2\theta } \left(\frac{1}{-1+\theta }\right)^{-2 \theta }}\times \nonumber \\
&&\left\{\alpha  (-2+\theta ) (-1+\theta )+3^{-2^{-\theta } \left(\frac{1}{-1+\theta }\right)^{\theta
}} B \left(\frac{K M^2}{B^2 \varepsilon  \theta ^2}\right)^{2^{-\theta } \left(\frac{1}{-1+\theta }\right)^{\theta }} \theta \times \right. \nonumber \\
&&\left. \left[\gamma  (-1+\theta)+3^{-2^{-\theta } \left(\frac{1}{-1+\theta }\right)^{\theta }} B \beta  \left(\frac{K M^2}{B^2 \varepsilon  \theta ^2}\right)^{2^{-\theta } \left(\frac{1}{-1+\theta}\right)^{\theta }} \theta \right]\right\},  \\
B_3 &=&B^2 \varepsilon  \theta ^2.
\end{eqnarray}
The solution of Eq. (\ref{81}) is given by
\begin{eqnarray}
F\left( K \right) = \frac{A_4}{B_4 }+C_8\sqrt{K}, \label{82}
\end{eqnarray}
where $C_8$ represents an integration constant, and
\begin{eqnarray}
A_4 &=& 2\ 3^{-4^{\theta } \left(\frac{1}{-1+\theta }\right)^{-2 \theta }-2^{1-\theta } \left(\frac{1}{-1+\theta
}\right)^{\theta }} c^2 K \left(\frac{K M^2}{B^2 \varepsilon  \theta ^2}\right)^{4^{\theta } \left(\frac{1}{-1+\theta }\right)^{-2 \theta }}\times \nonumber \\
&&\left\{\frac{2^{\theta} \beta  \left(\frac{1}{-1+\theta }\right)^{2 \theta } \left(\frac{K M^2}{B^2 \varepsilon  \theta ^2}\right)^{2^{1-\theta } \left(\frac{1}{-1+\theta}\right)^{\theta }}}{2^{1+3 \theta }+2^{\theta } \left(\frac{1}{-1+\theta }\right)^{2 \theta }+4 \left(\frac{1}{-1+\theta }\right)^{3 \theta }} \right. \nonumber \\
&&\left.
+\frac{2^{\theta} 3^{2^{-\theta } \left(\frac{1}{-1+\theta }\right)^{\theta }} \gamma  \left(\frac{1}{-1+\theta }\right)^{-1+2 \theta } \left(\frac{K M^2}{B^2 \varepsilon \theta ^2}\right)^{2^{-\theta } \left(\frac{1}{-1+\theta }\right)^{\theta }}}{2^{1+3 \theta } B \theta +2^{\theta } B \left(\frac{1}{-1+\theta }\right)^{2 \theta } \theta +2 B \left(\frac{1}{-1+\theta }\right)^{3 \theta } \theta } \right. \nonumber \\
&&\left.+\frac{3^{2^{1-\theta } \left(\frac{1}{-1+\theta }\right)^{\theta }} \alpha  \left(2-3 \theta +\theta ^2\right)}{B^2 \left[1+2^{1+2 \theta } \left(\frac{1}{-1+\theta }\right)^{-2 \theta }\right] \theta ^2}\right\},
 \\
B_4 &=&\varepsilon.
\end{eqnarray}
Using in Eq. (\ref{19}) the expression of $\rho_{EA}$ derived in Eq. (\ref{80}) along with the expression of $H$ obtained in Eq. (\ref{70}),
we can write the pressure $p_{EA}$ as
\begin{eqnarray}
p_{EA} = -\frac{A_5}{B_5}, \label{83}
\end{eqnarray}
where
\begin{eqnarray}
A_5 &=&  c^2 t^{-2-\theta } \left\{\alpha  (-2+\theta ) (-1+\theta ) \left(-2+3 B t^{\theta } \theta \right) \right. \nonumber \\
&&\left. +B t^{\theta } \theta  \left[\gamma (-1+\theta ) \left(-2+\theta +3 B t^{\theta } \theta \right)  +B t^{\theta } \beta  \theta  \left(-2+2 \theta +3 B t^{\theta } \theta \right)\right]\right\},  \\
B_5 &=& B \theta.
\end{eqnarray}
Therefore, the EoS parameter $\omega_{EA}$ is given by
\begin{eqnarray}
\omega_{EA} &=&  \frac{p_{EA}}{\rho_{EA}} =-1+\frac{2 t^{-\theta }}{3 B \theta }\nonumber \\
&&-\frac{\theta  \left(\gamma  (-1+\theta )+2 B t^{\theta } \beta  \theta \right)}{3 \left\{\alpha (-2+\theta ) (-1+\theta )+B t^{\theta } \theta  \left[\gamma  \left(-1+\theta \right)+B t^{\theta } \beta  \theta \right]\right\}}. \label{84}
\end{eqnarray}

It is possible to analyse   matter dominated universe  and the accelerated phase of the universe using a single formalism.
Now for such a model,  the Hubble parameter $H$ is  given by \cite{hep-th/0506212, ycyc}
\begin{equation}\label{hubbleassume}
H(t)=H_0+\frac{H_1}{t},
\end{equation}
with $H_0$ and $H_1$ being two constant parameters.
From Eq. (\ref{hubbleassume}), we can easily obtain the following expression of the scale factor $a(t)$
  \begin{equation}
  a(t)=C_9 e^{H_0 t}t^{H_1},
\end{equation}
where $C_9$ is an integration constant.
Moreover, using Eq. (\ref{hubbleassume}), we have that the first and the second time derivatives of the Hubble parameter,
by the following relations:
\begin{eqnarray}
\dot{H} &=&  -\frac{H_1}{t^2},  \label{71DD}\\
\ddot{H} &=&  \frac{2 H_1}{t^3}. \label{72DD}
\end{eqnarray}
Using in Eq. (\ref{13}) the expression of $H$ given in Eq. (\ref{hubbleassume}), we obtain the following expression for $K$,
\begin{eqnarray}
K = \frac{3 \varepsilon \left(H_0+\frac{H_1}{t}\right)^2}{M^2} \label{cappaDD}.
\end{eqnarray}
Using in the general expression of $L_{GO}$ given in Eq. (\ref{deflgo}),
the expressions of $H$ and $\dot{H}$ obtained in Eqs. (\ref{71DD}) and (\ref{72DD}),
we obtain
\begin{eqnarray}
L_{GO} =  \frac{1}{\sqrt{\frac{-\beta  H_1+\alpha  \left(t H_0+H_1\right)^2}{t^2}}} = \frac{t}{\sqrt{-\beta  H_1+\alpha  \left(t H_0+H_1\right)^2}}. \label{lgoDD}
\end{eqnarray}
Therefore, we can conclude that the expression of $\rho_{EA}$ with Granda-Oliveros cut-off can be written as
\begin{eqnarray}
\rho_{EA} = \frac{3 c^2 \left[-\beta  H_1+\alpha  \left(t H_0+H_1\right)^2\right]}{t^2} \label{mama1}.
\end{eqnarray}
Using in Eq. (\ref{18a}) the expressions of $H$ and $\rho_{EA}$ given in Eqs. (\ref{71DD}) and (\ref{mama1}),  or
equivalently, in Eq. (\ref{18b}) the expressions of $K$ and $\rho_{EA}$ given in Eqs. (\ref{cappaDD}) and (\ref{rhoDD}),
we obtain the following differential equation for $F\left( K\right)$
\begin{eqnarray}
\frac{dF}{dK}-\frac{F}{2 K}-\frac{3 c^2 \left[\frac{K M^2 \alpha }{3 \varepsilon}-\frac{\beta  \left(\frac{1}{3} \varepsilon K M^2-H_0\right)^2}{H_1}\right]}{KM^2}=0,
\end{eqnarray}
which solution is given by
\begin{eqnarray}
F\left(K\right) &=&  \sqrt{K} C_{10}\nonumber \\
&&-\frac{2 c^2 \left[-18 \varepsilon^2 K M^2 \beta  H_0-27 \varepsilon \beta  H_0^2+K M^2
\left(\varepsilon^3 K M^2 \beta -9 \alpha  H_1\right)\right]}{9 \varepsilon M^2 H_1}, \label{FDD}
\end{eqnarray}
where $C_{10}$ represents an integration constant.

Using in Eq. (\ref{19}) the expression of $\rho_{EA}$ derived in Eq. (\ref{mama1}) along with the expression of $H$ defined in  Eq. (\ref{hubbleassume}),
we can write the pressure $p_{EA}$ as
\begin{eqnarray}
p_{EA}= \frac{c^2 \left\{-2 \beta  H_1-\left(t H_0+H_1\right) \left[3 t^2 \alpha  H_0^2+\left(-2 \alpha -3 \beta +6 t \alpha  H_0\right) H_1+3\alpha  H_1^2\right]\right\}}{t^2 \left(t H_0+H_1\right)}.
\end{eqnarray}
Therefore, we have that the EoS parameter $\omega_{EA}$ for this case is given by
\begin{eqnarray}
 \omega_{EA}=-1+\frac{2}{3 \left(t H_0+H_1\right)}-\frac{2 t \alpha  H_0}{3 \left[-\beta  H_1+\alpha  \left(t H_0+H_1\right)^2\right]}.
\end{eqnarray}

At Ricci scale, i.e., for $\alpha =2$ and $\beta =1$, we obtain
\begin{eqnarray}
L_{GO} &=&  \frac{1}{\sqrt{\frac{-  H_1+2  \left(t H_0+H_1\right)^2}{t^2}}} = \frac{t}{\sqrt{-  H_1+2  \left(t H_0+H_1\right)^2}}, \label{lgoDD} \\
\rho_{EA} &=& \frac{3 c^2 \left[-  H_1+2  \left(t H_0+H_1\right)^2\right]}{t^2} \label{rhoDD},\\
F(K) &=&  -\frac{2 c^2 \left[- 18\varepsilon^2 K M^2   H_0-27 \varepsilon   H_0^2+K M^2
\left(\varepsilon^3 K M^2  -18  H_1\right)\right]}{9 \varepsilon M^2 H_1}\nonumber \\
&&+\sqrt{K} C_{10}, \label{FDD}\\
p_{EA}&=& \frac{c^2 \left\{-2   H_1-\left(t H_0+H_1\right) \left[3 t^2 \alpha  H_0^2+\left(-7  +12 t   H_0\right) H_1+6  H_1^2\right]\right\}}{t^2 \left(t H_0+H_1\right)},\\
 \omega_{EA}&=&-1+\frac{2}{3 \left(t H_0+H_1\right)}-\frac{4 t   H_0}{3 \left[-  H_1+2  \left(t H_0+H_1\right)^2\right]}.
\end{eqnarray}
For $\alpha \approx 0.8824$ and $\beta \approx 0.5016$, i.e., for the value of $\alpha$ and $\beta$ corresponding to a non-flat universe, we obtain
\begin{eqnarray}
L_{GO} &=&  \frac{1}{\sqrt{\frac{-0.5016  H_1+0.8824  \left(t H_0+H_1\right)^2}{t^2}}} = \frac{t}{\sqrt{-0.5016  H_1+0.8824  \left(t H_0+H_1\right)^2}}, \label{lgoDD} \\
\rho_{EA} &=& \frac{3 c^2 \left[-0.5016  H_1+0.8824  \left(t H_0+H_1\right)^2\right]}{t^2} \label{rhoDDD},\\
F\left(K\right) &=&  -\frac{2 c^2 \left[-9.0288 \varepsilon^2 K M^2   H_0- 13.5431\varepsilon   H_0^2+K M^2
\left(0.5016\varepsilon^3 K M^2  -7.9416  H_1\right)\right]}{9 \varepsilon M^2 H_1} \nonumber \\
&&+\sqrt{K} C_{10}, \label{FDD}\\
p_{EA}&=& \frac{c^2 }{t^2 \left(t H_0+H_1\right)}\times \left\{-1.0032  H_1-\left(t H_0+H_1\right) \times \right. \nonumber \\
&&\left.  \left[ 2.6472t^2  H_0^2+\left(-3.2696 +5.2944 t   H_0\right) H_1+   2.6472  H_1^2\right]\right\},\\
 \omega_{EA}&=&-1+\frac{2}{3 \left(t H_0+H_1\right)}-\frac{1.7648 t   H_0}{3 \left[-0.5016  H_1+0.8824  \left(t H_0+H_1\right)^2\right]}.
\end{eqnarray}
Furthermore, for $\alpha \approx 0.8502$ and $\beta\approx 0.4817$, i.e., for the value of $\alpha$ and $\beta$ corresponding to a  flat universe, we obtain
\begin{eqnarray}
L_{GO} &=&  \frac{1}{\sqrt{\frac{-0.4817  H_1+0.8502  \left(t H_0+H_1\right)^2}{t^2}}} = \frac{t}{\sqrt{-0.4817  H_1+0.8502  \left(t H_0+H_1\right)^2}}, \label{lgoDD} \\
\rho_{EA} &=& \frac{3 c^2 \left[-0.4817  H_1+0.8502  \left(t H_0+H_1\right)^2\right]}{t^2} \label{rhoDDminus},\\
F\left(K\right) &=& -\frac{2 c^2 \left[-8.6706 \varepsilon^2 K M^2   H_0-13.0059 \varepsilon  H_0^2+K M^2 \left(0.4817\varepsilon^3 K M^2  -7.5618  H_1\right)\right]}{9 \varepsilon M^2 H_1} \nonumber \\
&&+\sqrt{K} C_{10}, \label{FDD}\\
p_{EA}&=& \frac{c^2}{t^2 \left(t H_0+H_1\right)}\times \left\{-0.9634  H_1-\left(t H_0+H_1\right)\times \right. \nonumber \\
&&\left. \left[2.5506 t^2   H_0^2+\left(-3.1455 +5.1012 t   H_0\right) H_1+   2.5506 H_1^2\right]\right\},\\
\omega_{EA}&=&-1+\frac{2}{3 \left(t H_0+H_1\right)}-\frac{1.7004 t   H_0}{3 \left[-0.4817  H_1+0.8502  \left(t H_0+H_1\right)^2\right]}.
\end{eqnarray}

We now consider the Chen-Jing model studied in this paper, i.e.,  the energy density with higher derivatives of the Hubble parameter.
Using in Eq. (\ref{rho}) the expressions of $H$, $\dot{H}$ and $\ddot{H}$ given  in Eqs. (\ref{hubbleassume}), (\ref{71DD}) and (\ref{72DD}),
we obtain that the expression of $\rho_{EA}$,
\begin{eqnarray}
\rho_{EA}=  \frac{c^2 \left[-\gamma  H_1+\frac{2 \alpha  H_1+\beta  \left(t H_0+H_1\right){}^3}{t H_0+H_1}\right]}{3 \varepsilon t^2 \left(H_0+\frac{H_1}{t}\right)^2}. \label{casaleggio}
\end{eqnarray}
Following the same procedure as in  the previous case, we obtain a  differential equation for $F\left( K\right)$,
\begin{eqnarray}
F(K)&=&\sqrt{K}\left[C_{11}+\,{\frac {2{c}^{2}{M}^{2}\alpha\,{K}^{3/2}}{\epsilon\,{H_{{1}}}^{2}}}-
{\frac {{c}^{2}{M}^{2}\beta\,{K}^{3/2}}{\epsilon\,H_{{1}}}}+{\frac {{c
}^{2}{M}^{2}\gamma\,{K}^{3/2}}{\epsilon}}-\,{\frac {9{c}^{2}\sqrt {2}M
H_{{0}}\alpha\,K}{\sqrt {\epsilon}{H_{{1}}}^{2}}} \right. \nonumber \\
&&\left.+\,{\frac {3{c}^{2}
\sqrt {2}MH_{{0}}\beta\,K}{\sqrt {\epsilon}H_{{1}}}}+\,{\frac {
18\sqrt {2\epsilon}{c}^{2}{H_{{0}}}^{2}H\alpha\,\ln  \left( K
 \right) }{{H_{{1}}}^{2}M}}-\,{\frac {3\sqrt {2\epsilon}{c}^{2}{H_{{0}}}^{2}H\beta\,\ln  \left( K \right) }{H_{{1}}M}} \right.\nonumber \\
&&\left.-\,{\frac {
6\sqrt {2\epsilon}{c}^{2}\alpha\,{H_{{0}}}^{3}\ln  \left( K
 \right) }{{H_{{1}}}^{2}M}}
\right], \label{maurobiglino1}
\end{eqnarray}
where $C_{11}$  is an integration constant.
Using in Eq. (\ref{19}) the expression of $\rho_{EA}$ derived in Eq. (\ref{casaleggio}) along with the expression of $H$ obtained in Eq. (\ref{hubbleassume}),
we can write the pressure $p_{EA}$ as
\begin{eqnarray}
p_{EA}&=& \left\{c^2 \left[-6 \alpha -2 \gamma -3 t^2 \beta  H_0^2+\left(2 \beta +3 \gamma -6 t \beta  H_0\right) H_1-\right. \right. \nonumber \\
&& \left. \left.3 \beta  H_1^2-\frac{2 t^2\alpha  H_0^2}{\left(t H_0+H_1\right)^3}-\frac{2 t \alpha  H_0}{\left(t H_0+H_1\right){}^2}+\frac{4 \alpha +2 t (3 \alpha +\gamma ) H_0}{t H_0+H_1}\right]\right\}\times t^{-2}.
\end{eqnarray}
Therefore, we have that the EoS parameter $\omega_{EA}$ for this case is given by
\begin{eqnarray}
 \omega_{EA} &=& \left\{-6 \alpha +2 \gamma +3 t^2 \beta  H_0^2+\left(-2 \beta -3 \gamma +6 t \beta  H_0\right) H_1 \right.\nonumber \\
&&\left. +3 \beta  H_1^2+\frac{2 t^2 \alpha  H_0^2}{\left(t H_0+H_1\right){}^3}+\frac{2 t \alpha  H_0}{\left(t H_0+H_1\right){}^2}-\frac{2 \left[2 \alpha +t (3 \alpha +\gamma ) H_0\right]}{t H_0+H_1}\right\} \nonumber \\
&&\times\left[3 \left(-\gamma H_1+\frac{2 \alpha  H_1+\beta  \left(t H_0+H_1\right){}^3}{t H_0+H_1}\right)\right].
\end{eqnarray}

We can also analyse a $q$-de Sitter model  \cite{Setare:2014vna}.
The scale factor for such a model is given by
\begin{eqnarray}
&&a(t)=e_{q}(H_0 t)=\Big[1+(q-1)H_0 t\Big]^{\frac{1}{q-1}}\label{a(t)}.
\end{eqnarray}
 With the choice of scale factor,  we can derive that the Hubble parameter $H$ along with its first and second time
derivatives are given,
\begin{eqnarray}
H &=& H_0 \left[1+H_0 (-1+q) t\right]^{-1+\frac{1}{-1+q}},  \label{desitH}\\
\dot{H} &=& H_0^2 \left(-1+\frac{1}{-1+q}\right) (-1+q) \left[1+H_0 (-1+q) t\right]^{-2+\frac{1}{-1+q}},  \label{desitHdot}\\
\ddot{H} &=& H_0^3 \left(-2+\frac{1}{-1+q}\right) \left(-1+\frac{1}{-1+q}\right) \left(-1+q\right)^2\times \nonumber \\
&& \left[1+H_0 (-1+q) t\right]^{-3+\frac{1}{-1+q}}. \label{desitHddot}
\end{eqnarray}
Using in Eq. (\ref{13}) the expression of $H$ given in Eq. (\ref{desitH}), we obtain the following expression for $K$,
\begin{eqnarray}
 K= \frac{3 \varepsilon H_0^2 [1+H_0 (-1+q) t]^{-2+\frac{2}{-1+q}}}{M^2} \label{biglino2}.
\end{eqnarray}
Using in the general expression of $L_{GO}$ given in Eq. (\ref{deflgo}),
the expressions of $H$ and $\dot{H}$ obtained in Eqs. (\ref{desitH}) and (\ref{desitHdot}), we obtain
\begin{eqnarray}
L_{GO}= \frac{1}{\sqrt{H_0^2 [1+H_0 (-1+q) t]^{-2+\frac{1}{-1+q}} \left\{\left[1+H_0 (-1+q) t\right]^{\frac{1}{-1+q}} \alpha -(-2+q) \beta\right\}}}.
\end{eqnarray}
Therefore, we   conclude that the expression of $\rho_{EA}$ with Granda-Oliveros cut-off can be written as
\begin{eqnarray}
\rho_{EA}= 3 c^2 H_0^2 \left[1+H_0 (-1+q) t\right]^{-2+\frac{1}{-1+q}} \left\{\left[1+H_0 (-1+q) t\right]^{\frac{1}{-1+q}} \alpha -(-2+q) \beta \right\}.\label{grillino}
\end{eqnarray}
Using in Eq. (\ref{18a}) the expressions of $H$ and $\rho_{EA}$ given in Eqs. (\ref{desitH}) and (\ref{grillino}),  or
equivalently, in Eq. (\ref{18b}) the expressions of $K$ and $\rho_{EA}$ given in Eqs. (\ref{biglino2}) and (\ref{grillino}),
we a differential equation for $F\left( K\right)$ which solution is given by
\begin{eqnarray}
F(K) &=& 6 c^2 H_0^2 \left(\frac{K}{\varepsilon H_0^2}\right)^{1+\frac{1}{-2+q}} \left[\left(\frac{K}{\varepsilon H_0^2}\right)^{\frac{1}{-2+\frac{2}{-1+q}}}\right]^{\frac{1}{-1+q}} \times \nonumber \\
&&
\left[e^{\frac{q \left((-2+q) \log[K]+\log\left[\frac{K}{\varepsilon H_0^2}\right]+2 (-2+q) \log\left[\left(\frac{K}{\varepsilon
H_0^2}\right)^{-\frac{-1+q}{2 (-2+q)}}\right]\right)}{2 (-2+q)^2 (-1+q)}} K^{-\frac{q}{4-6 q+2 q^2}}\times \right.\nonumber \\
 && \left. \left(\frac{K}{\varepsilon H_0^2}\right)^{-\frac{-1+2q}{2
(-2+q)^2 (-1+q)}} \alpha -\frac{(-2+q)^2
\beta }{-1+q}\right]   \times M^{-2} + \sqrt{K}C_{12}, \label{aldoolanda1}
\end{eqnarray}
where $C_{12}$ is a constant parameter.

Using in Eq. (\ref{19}) the expression of $\rho_{EA}$ derived in Eq. (\ref{grillino}) along with the expression of $H$ defined in  Eq. (\ref{desitH}),
we can write the pressure $p_{EA}$ as
\begin{eqnarray}
p_{EA}&=& \frac{c^2 H_0^2}{\left[1+H_0 (-1+q) t\right]^2} \times\left\{-3 \left[1+H_0 (-1+q) t\right]^{\frac{2}{-1+q}} \alpha \right. \nonumber \\
&& \left.+(-6+(7-2 q) q) \beta +(-2+q)\left[1+H_0 (-1+q) t\right]^{\frac{1}{-1+q}} (2 \alpha +3 \beta )\right\}.
\end{eqnarray}
Therefore,   the EoS parameter $\omega_{EA}$ for this case is given by
\begin{eqnarray}
\omega_{EA}&=& \left\{-3 \alpha +\left[-6+(7-2 q) q\right] \left[1+H_0 (-1+q) t\right]^{-\frac{2}{-1+q}} \beta \right. \nonumber \\
&&\left.+(-2+q) \left[1+H_0 (-1+q) t\right]^{\frac{1}{1-q}} (2 \alpha+3 \beta )\right\} \times  \nonumber \\
&&\left\{3 \alpha -3 (-2+q) (1+H_0 (-1+q) t)^{\frac{1}{1-q}} \beta \right\}^{-1} .
\end{eqnarray}
At Ricci scale, i.e., for $\alpha =2$ and $\beta =1$, we obtain
\begin{eqnarray}
L_{GO}&=& \left\{H_0^2 [1+H_0 \left(-1+q\right) t]^{-2+\frac{1}{-1+q}} \times \right. \nonumber \\
&&\left.\left\{\left[1+H_0 (-1+q) t\right]^{\frac{1}{-1+q}} 2 -(-2+q) \right\}\right\}^{-1/2}, \\
\rho_{EA}&=& 3 c^2 H_0^2 \left[1+H_0 (-1+q) t\right]^{-2+\frac{1}{-1+q}}\times \nonumber \\
 &&\left\{\left[1+H_0 (-1+q) t\right]^{\frac{1}{-1+q}} 2 -(-2+q)  \right\},\\
F\left(K\right) &=& 6 c^2 H_0^2 \left(\frac{K}{\varepsilon H_0^2}\right)^{1+\frac{1}{-2+q}} \left[\left(\frac{K}{\varepsilon H_0^2}\right)^{\frac{1}{-2+\frac{2}{-1+q}}}\right]^{\frac{1}{-1+q}} \times \nonumber \\&& \left[e^{\frac{q \left((-2+q) \log[K]+\log\left[\frac{K}{\varepsilon H_0^2}\right]+2 (-2+q) \log\left[\left(\frac{K}{\varepsilon
H_0^2}\right)^{-\frac{-1+q}{2 (-2+q)}}\right]\right)}{2 (-2+q)^2 (-1+q)}} K^{-\frac{q}{4-6 q+2 q^2}}\times \right.\nonumber \\
 && \left. \left(\frac{K}{\varepsilon H_0^2}\right)^{-\frac{-1+2q}{2 (-2+q)^2 (-1+q)}} 2 -\frac{(-2+q)^2  }{-1+q}\right]   \times M^{-2} + \sqrt{K}C_{11},\\
p_{EA}&=& \frac{c^2 H_0^2}{\left[1+H_0 (-1+q) t\right]^2} \times\left\{-3 \left[1+H_0 (-1+q) t\right]^{\frac{2}{-1+q}} 2 \right. \nonumber \\
&& \left.+(-6+(7-2 q) q)  +(-2+q)\left[1+H_0 (-1+q) t\right]^{\frac{1}{-1+q}}\cdot 7\right\},\\
\omega_{EA}&=& \left\{-6 +\left[-6+(7-2 q) q\right] \left[1+H_0 (-1+q) t\right]^{-\frac{2}{-1+q}}  \right. \nonumber \\
&&\left.+(-2+q) \left[1+H_0 (-1+q) t\right]^{\frac{1}{1-q}} \cdot 7 \right\} \times  \nonumber \\
&&\left\{6 -3 (-2+q) (1+H_0 (-1+q) t)^{\frac{1}{1-q}}  \right\}^{-1} .
\end{eqnarray}
For $\alpha \approx 0.8824$ and $\beta \approx 0.5016$, i.e., for the value of $\alpha$ and $\beta$ corresponding to a non-flat universe, we obtain
\begin{eqnarray}
L_{GO}&=& \left\{H_0^2 [1+H_0 (-1+q) t]^{-2+\frac{1}{-1+q}}\times \right. \nonumber \\
 &&\left. \left\{\left[1+H_0 (-1+q) t\right]^{\frac{1}{-1+q}} 0.8824 -0.5016(-2+q) \right\}\right\}^{-1/2}, \\
\rho_{EA}&=& 3 c^2 H_0^2 \left[1+H_0 (-1+q) t\right]^{-2+\frac{1}{-1+q}}\times \nonumber \\
&& \left\{\left[1+H_0 (-1+q) t\right]^{\frac{1}{-1+q}} 0.8824 -0.5016(-2+q) \right\},\\
F(K) &=& 6 c^2 H_0^2 \left(\frac{K}{\varepsilon H_0^2}\right)^{1+\frac{1}{-2+q}} \left[\left(\frac{K}{\varepsilon H_0^2}\right)^{\frac{1}{-2+\frac{2}{-1+q}}}\right]^{\frac{1}{-1+q}} \times \nonumber \\&& \left[e^{\frac{q \left((-2+q) \log[K]+\log\left[\frac{K}{\varepsilon H_0^2}\right]+2 (-2+q) \log\left[\left(\frac{K}{\varepsilon
H_0^2}\right)^{-\frac{-1+q}{2 (-2+q)}}\right]\right)}{2 (-2+q)^2 (-1+q)}} K^{-\frac{q}{4-6 q+2 q^2}}\times \right.\nonumber \\
 && \left. \left(\frac{K}{\varepsilon H_0^2}\right)^{-\frac{-1+2q}{2 (-2+q)^2 (-1+q)}} 0.8824 -\frac{(-2+q)^2 \cdot 0.5016 }{-1+q}\right]   M^{-2} + \sqrt{K}C_{11},\\
p_{EA}&=& \frac{c^2 H_0^2}{\left[1+H_0 (-1+q) t\right]^2} \times\left\{-3 \left[1+H_0 (-1+q) t\right]^{\frac{2}{-1+q}} 0.8824 \right. \nonumber \\
&& \left.+\left[\left(7-2 q\right) q-6\right] \cdot 0.5016 +(-2+q)\left[1+H_0 (-1+q) t\right]^{\frac{1}{-1+q}}\cdot 3.2696 \right\},\\
\omega_{EA}&=& \left\{-2.6472 +\left[-6+(7-2 q) q\right] \left[1+H_0 (-1+q) t\right]^{-\frac{2}{-1+q}} \cdot 0.5016 \right. \nonumber \\
&&\left.+(-2+q) \left[1+H_0 (-1+q) t\right]^{\frac{1}{1-q}} \cdot 3.2696\right\} \times  \nonumber \\
&&\left\{2.6472 -3 (-2+q) (1+H_0 (-1+q) t)^{\frac{1}{1-q}} \cdot 0.5016 \right\}^{-1} .
\end{eqnarray}
Furthermore, for $\alpha \approx 0.8502$ and $\beta \approx 0.4817$, i.e., for the value of $\alpha$ and $\beta$ corresponding to a  flat Universe, we obtain
\begin{eqnarray}
L_{GO}&=& \left\{H_0^2 [1+H_0 (-1+q) t]^{-2+\frac{1}{-1+q}} \times \right.\nonumber \\
&&\left.\left\{\left[1+H_0 (-1+q) t\right]^{\frac{1}{-1+q}} 0.8502 -0.4817(-2+q) \right\}\right\}^{-1/2}, \\
\rho_{EA}&=& 3 c^2 H_0^2 \left[1+H_0 (-1+q) t\right]^{-2+\frac{1}{-1+q}}\times \nonumber \\
&& \left\{\left[1+H_0 (-1+q) t\right]^{\frac{1}{-1+q}} 0.8502 -0.4817(-2+q) \right\},\\
F(K) &=& 6 c^2 H_0^2 \left(\frac{K}{\varepsilon H_0^2}\right)^{1+\frac{1}{-2+q}} \left[\left(\frac{K}{\varepsilon H_0^2}\right)^{\frac{1}{-2+\frac{2}{-1+q}}}\right]^{\frac{1}{-1+q}} \times \nonumber \\&& \left[e^{\frac{q \left((-2+q) \log[K]+\log\left[\frac{K}{\varepsilon H_0^2}\right]+2 (-2+q) \log\left[\left(\frac{K}{\varepsilon
H_0^2}\right)^{-\frac{-1+q}{2 (-2+q)}}\right]\right)}{2 (-2+q)^2 (-1+q)}} K^{-\frac{q}{4-6 q+2 q^2}}\times \right.\nonumber \\
 && \left. \left(\frac{K}{\varepsilon H_0^2}\right)^{-\frac{-1+2q}{2 (-2+q)^2 (-1+q)}} 0.8502 -\frac{(-2+q)^2 \cdot 0.4817 }{-1+q}\right]    M^{-2} + \sqrt{K}C_{11},\\
p_{EA}&=& \frac{c^2 H_0^2}{\left[1+H_0 (-1+q) t\right]^2} \times\left\{-3 \left[1+H_0 (-1+q) t\right]^{\frac{2}{-1+q}} 0.8502 \right. \nonumber \\
&& \left.+\left[\left(7-2 q\right) q-6\right] \cdot 0.4817 +(-2+q)\left[1+H_0 (-1+q) t\right]^{\frac{1}{-1+q}}\cdot 3.1455 \right\},\\
\omega_{EA}&=& \left\{-2.5506 +\left[-6+(7-2 q) q\right] \left[1+H_0 (-1+q) t\right]^{-\frac{2}{-1+q}} 0.4817 \right. \nonumber \\
&&\left.+(-2+q) \left[1+H_0 (-1+q) t\right]^{\frac{1}{1-q}}\cdot 3.1455\right\} \times  \nonumber \\
&&\left\{2.5506 -3 (-2+q) (1+H_0 (-1+q) t)^{\frac{1}{1-q}} 0.4817 \right\}^{-1} .
\end{eqnarray}

We now consider the Chen-Jing model   i.e., the one with the first and the second time derivatives of the Hubble parameter $H$.
Using in Eq. (\ref{rho}) the expressions of $H$, $\dot{H}$ and $\ddot{H}$ obtained in Eqs. (\ref{desitH}), (\ref{desitHdot}) and (\ref{desitHdot}),
we obtain that the expression of $\rho_{EA}$
\begin{eqnarray}
\rho_{EA} &=& 3 c^2 H_0^2 \left\{(-2+q) (-3+2 q) \alpha +\left[1+H_0 (-1+q) t\right]^{\frac{1}{-1+q}} \times \right. \nonumber \\
&&\left.\left[(1+H_0 (-1+q) t)^{\frac{1}{-1+q}} \beta -(-2+q) \gamma \right]\right\}   \times \left[1+H_0 (-1+q) t\right]^{-2}. \label{cate1}
\end{eqnarray}
Following the same procedure as the previous case, we obtain a  differential equation for $F\left( K\right)$ whose solution is given by
\begin{eqnarray}
F(K)=\sqrt {K} C_{13}+\,{\frac {3{c}^{2} \left( 2\,
\alpha\,{q}^{2}-4\,\alpha\,q+2\,\alpha-\beta\,q+\beta+\gamma \right) }
{\epsilon}}K,
\end{eqnarray}
where $C_{13}$ is a constant of integration.

Using in Eq. (\ref{19}) the expression of $\rho_{EA}$ derived in Eq. (\ref{cate1}) along with the expression of $H$ obtained in Eq. (\ref{desitH}),
we can write the pressure $p_{EA}$ as
\begin{eqnarray}
 p_{EA}&=&-c^2 H_0^2 (1+H_0 (-1+q) t)^{-2+\frac{1}{1-q}}\times \nonumber \\
 && \left\{-2 (-2+q) (-1+q) (-3+2 q) \alpha +3 \left[1+H_0 (-1+q) t\right]^{\frac{3}{-1+q}}
\beta \right. \nonumber \\
&&\left.+(-2+q) (-3+2 q) \left[1+H_0 (-1+q) t\right]^{\frac{1}{-1+q}} (3 \alpha +\gamma )-\right. \nonumber \\
&&\left.(-2+q) \left[1+H_0 (-1+q) t\right]^{\frac{2}{-1+q}} (2 \beta +3 \gamma
)\right\}.
\end{eqnarray}
Thus,    the EoS parameter $\omega_{EA}$ for this case is given by
\begin{eqnarray}
\omega_{EA}&=& -  (1+H_0 (-1+q) t)^{\frac{1}{1-q}}\times \nonumber \\
&& \left\{-2 (-2+q) (-1+q) (-3+2 q) \alpha +3 (1+H_0 (-1+q) t)^{\frac{3}{-1+q}} \beta \right. \nonumber \\
&&\left.+(-2+q) (-3+2 q) (1+H_0 (-1+q) t)^{\frac{1}{-1+q}} (3 \alpha +\gamma ) \right. \nonumber \\
&&\left.-(-2+q) (1+H_0 (-1+q) t)^{\frac{2}{-1+q}} (2 \beta +3 \gamma )\right\} \nonumber \\
&&\times \left\{3\left[(-2+q) (-3+2 q) \alpha +\left[1+H_0 (-1+q) t\right]^{\frac{1}{-1+q}}\times \right. \right. \nonumber \\
&&\left. \left. \left[(1+H_0 (-1+q) t)^{\frac{1}{-1+q}} \beta -(-2+q) \gamma \right]\right]\right\}^{-1}.
\end{eqnarray}

 \section{Appendix}
 In this appendix, we provide the $H(z)$ measurements (in unit [$\mathrm{km\,s^{-1}Mpc^{-1}}$]) and their errors \cite{Farooq:2013hq}.
\begin{table}
\caption{$H(z)$ measurements (in unit [$\mathrm{km\,s^{-1}Mpc^{-1}}$]) and their errors \cite{Farooq:2013hq}
.}
\begin{center}
\label{hubble}
\begin{tabular}{ccc}
\hline\hline
~$z$ & ~~$H(z)$ &~~ $\sigma_{H}$ \\
0.070&~~    69&~~~~~~~  19.6\\
0.100&~~    69&~~~~~~~  12\\
0.120&~~    68.6&~~~~~~~    26.2\\
0.170&~~    83&~~~~~~~  8\\
0.179&~~    75&~~~~~~~  4\\
0.199&~~    75&~~~~~~~  5\\
0.200&~~    72.9&~~~~~~~    29.6\\
0.270&~~    77&~~~~~~~  14\\
0.280&~~    88.8&~~~~~~~    36.6\\
0.350&~~    76.3&~~~~~~~    5.6\\
0.352&~~    83&~~~~~~~  14\\
0.400&~~    95&~~~~~~~  17\\
0.440&~~    82.6&~~~~~~~    7.8\\
0.480&~~    97&~~~~~~~  62\\
0.593&~~    104&~~~~~~~ 13\\
0.600&~~    87.9&~~~~~~~    6.1\\
0.680&~~    92&~~~~~~~  8\\
0.730&~~    97.3&~~~~~~~    7.0\\
0.781&~~    105&~~~~~~~ 12\\
0.875&~~    125&~~~~~~~ 17\\
0.880&~~    90&~~~~~~~  40\\
0.900&~~    117&~~~~~~~ 23\\
1.037&~~    154&~~~~~~~ 20\\
1.300&~~    168&~~~~~~~ 17\\
1.430&~~    177&~~~~~~~ 18\\
1.530&~~    140&~~~~~~~ 14\\
1.750&~~    202&~~~~~~~ 40\\
2.300&~~    224&~~~~~~~ 8\\

\hline\hline
\end{tabular}
\end{center}
\end{table}


\begin{thebibliography}{100}





\bibitem{Hooft} G. 't Hooft, {\it Quantization of Point Particles in 2+1 Dimensional Gravity and Space-Time Discreteness}, Class. Quantum Gravit. \textbf{13}, (1996) 1023 [ gr-qc/9601014]

\bibitem{Samuel1} V. A. Kostelecky and S. Samuel, {\it Spontaneous breaking of Lorentz symmetry in string theory}, Phys. Rev. D \textbf{39},(1989) 683 [IUHET-139, CCNY-HEP-88/4 ]

\bibitem{Ellis} G. Amelino-Camelia, J. R. Ellis, N. Mavromatos, D. V. Nanopoulos and S. Sarkar, {\it Tests of quantum gravity from observations of big gamma-ray bursts}, Nature \textbf{393},(1998) 763 [astro-ph/9712103]

\bibitem{Gambini} R. Gambini and J. Pullin, {\it Nonstandard optics from quantum space-time}, Phys. Rev. D \textbf{59},(1999) 124021[ gr-qc/9809038].

\bibitem{FaizalMPLA} M. Faizal, {\it Noncommutativity and Non-Anticommutativity Perturbative Quantum Gravity},  Mod. Phys. Lett. A \textbf{27}, (2012) 1250075 [arXiv:1204.0295].

\bibitem{Carroll} S.~M. Carroll, J.~A. Harvey, V.~A. Kostelecky, C.~D. Lane and T.~Okamoto, {\it Noncommutative Field Theory and Lorentz Violation}, Phys. Rev. Lett. \textbf{87},(2001) 141601 [ hep-th/0105082].

\bibitem{FaizalJPA} M. Faizal, {\it Spontaneous breaking of Lorentz symmetry by ghost condensation in perturbative quantum gravity}, J. Phys. A \textbf{44}, (2011) 402001 [ arXiv:1108.2853].

\bibitem{Greisen} K. Greisen,{\it End to the Cosmic-Ray Spectrum?}, Phys. Rev. Lett. \textbf{16},(1966) 748 [DOI: 10.1103/PhysRevLett.16.748].

\bibitem{Zatsepin} G. T. Zatsepin and V. A. Kuzmin, {\it Upper Limit of the Spectrum of Cosmic Rays}, JETP Lett. \textbf{4},(1966) 78[Pisma Zh.Eksp.Teor.Fiz. 4 (1966) 114-117].

\bibitem{Abraham} J. Abraham et al. (Pierre Auger Collaboration), 	{\it Measurement of the energy spectrum of cosmic rays above 1018 eV using the Pierre Auger Observatory},  Phys. Lett. B \textbf{685}, (2010) 239[arXiv:1002.197].

\bibitem{HoravaPRD} P.~Horava, {\it Quantum gravity at a Lifshitz point}, Phys. Rev. D \textbf{79}, 084008 (2009)[ arXiv:0901.3775].

\bibitem{HoravaPRL} P.~Horava, {\it Spectral dimension of the universe in quantum gravity at a Lifshitz point}, Phys. Rev. Lett. \textbf{102}, (2009) 161301 [arXiv:0902.3657]

\bibitem{1} R. G. Cai, B. Hu and H. B. Zhang, {\it Dynamical Scalar Degree of Freedom in Horava-Lifshitz Gravity}, Phys. Rev. D 80, (2009) 041501 [arXiv:0905.0255].
\bibitem{1a} C. Charmousis, G. Niz, A. Padilla and P. M. Saffin, {\it Strong coupling in Horava gravity}, JHEP 0908,(2009) 070 [arXiv:0905.2579].
\bibitem{1b} M. Li and Y. Pang, {\it A trouble with Horava-Lifshitz gravity}, JHEP 0908, (2009) 015 [ arXiv:0905.2751].
\bibitem{1c}  T. P. Sotiriou, M. Visser and S. Weinfurtner, {\it Quantum gravity without Lorentz invariance}, JHEP 0910,(2009) 033 [arXiv:0905.2798].
\bibitem{d1}  D. Blas, O. Pujolas  and  S.  Sibiryakov, {\it On the Extra Mode and Inconsistency of Horava Gravity}, JHEP 0910, (2009) 029 [ arXiv:0906.3046].
\bibitem{1d}  A. A. Kocharyan, {\it Is nonrelativistic gravity possible?},  Phys. Rev. D 80, (2009) 024026 [ arXiv:0905.4204].
\bibitem{2} D. Blas, O. Pujolas and S. Sibiryakov, {\it Consistent Extension of Horava Gravity}, Phys. Rev. Lett. 104,(2010) 181302[ arXiv:0909.3525].
\bibitem{2a}  T.  Jacobson, {\it Extended Horava gravity and Einstein-aether theory}, Phys. Rev. D81,(2010) 101502[ arXiv:1001.4823 ].

\bibitem{4a}  T. Jacobson, {\it Einstein-aether gravity: a status report}, [arXiv:0801.1547]

\bibitem{bi} K. Yagi, D.  Blas, E. Barausse and  N.  Yunes,  {\it Constraints on Einstein-aether theory and Horava gravity from binary pulsar observations},	Phys. Rev. D 89,(2014) 084067 [ arXiv:1311.7144].

\bibitem{5a}  C.  Heinicke, P. Baekler and  F.  W. Hehl, {\it Einstein-aether theory, violation of Lorentz invariance, and metric-affine gravity}, Phys. Rev. D 72, (2005) 02501 [arXiv:gr-qc/0504005].

\bibitem{6a}  E. Barausse, T. Jacobson and  T. P. Sotiriou, {\it Black holes in Einstein-aether and Horava-Lifshitz gravity}, Phys. Rev. D 83, (2011) 124043[ arXiv:1104.2889].

\bibitem{as}  J.  D. Barrow, {\it Some inflationary Einstein-Aether cosmologies}, Phys. Rev. D 85,(2012) 047503[ arXiv:1201.288].
\bibitem{po}  H.  Wei, X. P.  Yan and  Y. N.  Zhou, {\it Cosmological Evolution of Einstein-Aether Models with Power-law-like Potential}, Gen. Rel. Grav. 46, (2014) 1719 [arXiv:1310.533].
\bibitem{ga}  Z.  Haghani, T.  Harko, H. R.  Sepangi and  S.  Shahidi, {\it Cosmology of a Lorentz violating Galileon theory}, JCAP 05,(2015) 022[ arXiv:1501.0081].
\bibitem{gw} I. D. Saltas, I. Sawicki,   L,  Amendola and  M.  Kunz, {\it Anisotropic stress as signature of non-standard propagation of gravitational waves}, Phys. Rev. Lett. 113,(2014) 191101[ arXiv:1406.7139].

\bibitem{48} T. G Zlosnik, P. G Ferreira and G. D Starkman, {\it Modifying gravity with the aether: An alternative to dark matter},  Phys. Rev. D 75, (2007) 044017 [ astro-ph/0607411].
\bibitem{49} T. G. Zlosnik, P.G.  Ferreira and G. D. Starkman, {\it  Growth of structure in theories with a dynamical preferred frame}, Phys. Rev. D 77,  084 (2008) 010 [arXiv:0711.0520 ].

\bibitem{data}S.  Perlmutter, et al., {\it Measurements of Omega and Lambda from 42 High-Redshift Supernovae},  Astrophys. J. 517, (1999) 565 [astro-ph/9812133].

\bibitem{data2} A. G. Riess   et al., {\it Observational Evidence from Supernovae for an Accelerating Universe and a Cosmological Constant},   Astron. J. 116, (1988) 1009 [ astro-ph/9805201].
\bibitem{data3} D. N. Spergel et al.,  [WMAP Collaboration], {\it First-Year Wilkinson Microwave Anisotropy Probe (WMAP) Observations: Determination of Cosmological Parameters},   Astrophys. J. Suppl. 148,(2003) 175 [ astro-ph/0302209].
\bibitem{data4}D. N.  Spergel  et al., [WMAP Collaboration], {\it Three-Year Wilkinson Microwave Anisotropy Probe
(WMAP) Observations: Implications for Cosmology},  Astrophys. J. Suppl. 170, (2007) 377 [astro-ph/0603449].

\bibitem{data5} E. Komatsu  et al.,  {\it Five-year Wilkinson Microwave Anisotropy Probe observations: Cosmological
interpretation},   Astrophys. J. Suppl. 180, (2009) 330 [ arXiv:0803.0547].

\bibitem{data6} E. Komatsu  et al., [WMAP Collaboration], {\it Seven-year Wilkinson Microwave Anisotropy Probe
(WMAP) Observations: Cosmological Interpretation},  Astrophys. J. Suppl. 192, (2011) 18 [ arXiv:1001.4538].

\bibitem{data7} M. Tegmark   et al., {\it Cosmological parameters from SDSS and WMAP}, Phys. Rev. D 69, (2004) 103501 [ astro-ph/0310723].
\bibitem{data8} U.  Seljak et al., {\it Cosmological parameter analysis including SDSS Lya forest and galaxy bias: Constraints on the primordial spectrum of fluctuations, neutrino mass, and dark energy},  Phys. Rev. D 71, (2005) 103515 [ astro-ph/0407372 ].
\bibitem{data9} D. J. Eisenstein  et al., {\it Detection of the Baryon Acoustic Peak in the Large-Scale Correlation
Function of SDSS Luminous Red Galaxies},  Astrophys. J.  633,(2005) 560 [ astro-ph/0501171].
\bibitem{data10} B. Jain   and A. Taylor, {\it Cross-Correlation Tomography: Measuring Dark Energy Evolution with
Weak Lensing}, Phys. Rev. Lett. 91, (2003) 141302 [ astro-ph/030604].






\bibitem{lgo} L. N.  Granda and A.  Oliveros, {\it New infrared cut-off for the holographic scalar fields models of dark
energy},  Phys. Lett. B, 671,(2009) 199 [ arXiv:0810.3663].


\bibitem{wangalfa} Y. Wang and L.  Xu, {\it Current observational constraints to the holographic dark energy model with a
new infrared cutoff via the Markov chain Monte Carlo method},  Phys. Rev. D 81, (2010) 083523 [ arXiv:1004.3340].


\bibitem{chens} S. Chen S, and J.  Jing, {\it Dark energy model with higher derivative of Hubble parameter}, Phys. Lett. B, 679,(2009) 144 [arXiv:0904.2950 ].
\bibitem{grandaoliverosa} L. N.  Granda and A. Oliveros, {\it Infrared cut-off proposal for the holographic density}, Phys. Lett. B, 669,(2008) 275 [arXiv:0810.3149].
\bibitem{grandaoliverosb} A. Khodam-Mohammadi, {\it Power-Law Entropy Corrected New Holographic Scalar Field Models
of Dark Energy with Modified Ir-Cutoff}, Mod. Phys. Lett. A 26, (2011) 2487 [arXiv:1107.5455].

\bibitem{Rani:2014sia} S. Rani, A. Altaibayeva, M. Shahalam, J. K.  Singh and R.  Myrzakulov, {\it Constraints on
cosmological parameters in power-law cosmology}, JCAP  1503,   03, (2015) 031 [arXiv:1407.3445].


\bibitem{power1} P. D.  Mannheim, {\it Conformal cosmology with no cosmological constant}, Gen. Rel. Grav.  22, (1990) 289.
\bibitem{power11} Allen R E, (1999),  {\it Four testable predictions of instanton cosmology},
  AIP Conf.\ Proc.\  {\bf 478}, 204 (1999)
  doi:10.1063/1.59392
  [astro-ph/9902042].

\bibitem{power2} K.  Bamba, A. N.  Makarenko, A. N.  Myagky  and S. D. Odintsov, {\it Bouncing cosmology in modified
Gauss-Bonnet gravity}, Phys. Lett. B   732, (2014) 349 [ arXiv:1403.3242].

\bibitem{power22}  K.  Bamba A. N.  Makarenko, A. N.  Myagky, S.  Nojiri   and S. D. Odintsov, {\it Bounce cosmology from
F(R) gravity and F(R) bigravity}, JCAP 01, (2014)008 [ arXiv:1309.3748].

\bibitem{a} R. Rangdee  and B.  Gumjudpai, {\it Tachyonic (phantom) power-law cosmology}, Astrophys. Space Science 349,(2014) 975 [arXiv:1210.5550].

\bibitem{za} S. Nojiri, S. D. Odintsov and S. Tsujikawa, {\it Properties of singularities in the (phantom) dark energy universe}, Phys. Rev. D 71, (2005) 063004 [ hep-th/0501025 ].
\bibitem{18} J. D. Barrow and A. R.  Liddle, {\it Perturbation spectra from intermediate inflation},  Phys. Rev. D47, (1993) 5219 [astro-ph/9303011].

\bibitem{31} P. B. Khatua and U.  Debnath, {\it Role of chameleon field in accelerating Universe},   Astrophys. Space Sci. 326, (2010) 53 [arXiv:1012.1443].

\bibitem{17} S.  Mukherjee, B. C.  Paul, N. K.  Dadhich, S. D.  Maharaj and A. Beesham, {\it Emergent universe with
exotic matter},  Class. Quant. Grav.  23, (2006) 6927 [ gr-qc/0605134].

\bibitem{30} B. C. Paul and S.  Ghose, {\it Emergent universe scenario in the Einstein-Gauss-Bonnet gravity with
dilaton}, Gen. Rel. Gravit. 42, (2010) 795 [ arXiv:0809.4131].






\bibitem{md} F. Darabi, {\it Acceleration of the Universe in Matter Dominant Era by Conformal Symmetry Breaking}, Int. J. Theor. Phys. 53, (2014) 881 [arXiv:1305.5378].

\bibitem{dm} S. Nesseris, S. Basilakos, E. N. Saridakis and  L. Perivolaropoulos, {\it Viable  f(T) models are practically indistinguishable from $\Lambda$CDM}, Phys. Rev. D 88,(2013) 103010 [ arXiv:1308.6142].

\bibitem{hep-th/0506212} S. Nojiri S and S. D. Odintsov, {\it Unifying phantom inflation with late-time acceleration: scalar
phantom-non-phantom transition model and generalized holographic dark energy}, Gen. Rel. Grav.  38,(2006) 1285 [ hep-th/0506212].

\bibitem{ycyc} S. Nojiri and S. D.  Odintsov, {\it Accelerating cosmology in modified gravity: from convenient {\it F(R)} or string-inspired theory to bimetric {\it F(R)} gravity}, Int. J. Geom. Meth. Mod. Phys. 11, (2014) 1460006 [arXiv:1306.4426].

\bibitem{qd}  D. A. Lowe, {\it q-deformed de Sitter/conformal field theory correspondence}, Phys. Rev. D 70,(2004) 104002 [hep-th/040718].


\bibitem{Setare:2014vna}   M. R. Setare, D.  Momeni, V.   Kamali and R. Myrzakulov, {\it Inflation driven by q-de Sitter}, Int. J. Theor. Phys. 55, (2016)1003 [arXiv:1409.3200].

\bibitem{Farooq:2013hq} O. Farooq  and B. Ratra, {\it Hubble Parameter Measurement Constraints on the Cosmological Deceleration-Acceleration Transition Redshift}, Astrophys. J.  766, (2013)L7 [ arXiv:1301.5243].

\bibitem{Planck} P. A. R. Ade,  et al., {\it Cosmological parameters}, [Planck Collaboration]   arXiv:1303.5076[astro-ph.CO].

\bibitem{sah}  V. Sahni, T. D.  Saini, A. A.   Starobinsky  and U. Alam, {\it Statefinder: A new geometrical diagnostic of
dark energy}, Soviet Journal of Experimental and Theoretical Physics Letters, 77,(2003) 201 [astro-ph/0201498].

\bibitem{alam} U. Alam, V. Sahni, T.  D.  Saini  and A. A.  Starobinsky, {\it Exploring the expanding Universe and
dark energy using the statefinder diagnostic},  Mon. Not. R. Astron. Soc., 344,(2003) 1057 [ astro-ph/0303009].

\bibitem{huang} Z. G. Huang, X. M.  Song, H. Q.  Lu   and W. Fang W, {\it Statefinder diagnostic for dilaton dark energy}, Astrophys. Space Sci.  315,(2008) 175 [arXiv:0802.2320].

\bibitem{wu1} P. Wu  and H. Yu H, {\it Observational constraints on f(T) theory},  Phys.  Lett.  B 693,(2010) 415 [arXiv:1006.0674].

\bibitem{wang} F. Y. Wang, Z. G. Dai and S. Qi, {\it Probing the cosmographic parameters to distinguish between dark
energy and modified gravity models},  Astron. Astrophys. 507,(2009) 53 [arXiv:0912.5141].

\bibitem{Sahni} V. Sahni, A. Shafieloo  and A. A. Starobinsky, {\it Two new diagnostics of dark energy}, Phys. Rev. D 78,(2008) 103502 [arXiv:0807.3548].

\bibitem{Shahalam:2015lra} M. Shahalam, S. Sami S and A. Agarwal, {\it  {\it Om} diagnostic applied to scalar field models and slowing down of cosmic acceleration}, Mon. Not. Roy. Astron. Soc. 448, (2015) 2948 [ arXiv:1501.04047].

\bibitem{Jamil:2013yc}  M. Jamil, D.  Momeni   and R. Myrzakulov,  {\it Observational constraints on non-minimally coupled Galileon model}, Eur. Phys. J. C  73,  (2013) 2347 [arXiv:1302.0129].

\bibitem{deFromont:2013iwa} P. de Fromont, C.  de Rham, L.  Heisenberg  and A. Matas, {\it Superluminality in the Bi- and Multi-Galileon}, JHEP 1307,(2013) 067 [arXiv:1303.0274].

\bibitem{nonc} W. Westra  and  S. Zohren, {\it A local induced action for the noncritical string}, Class. Quantum Grav. 29,(2012) 095021 [arXiv:1106.1460].

\bibitem{A12}  R. Gregory, S. L. Parameswaran, G. Tasinato and I. Zavala, {\it Lifshitz solutions in supergravity and string theory}, JHEP 1012,(2010) 047 [arXiv:1009.3445].
\bibitem{B12}  P.~Burda, R.~Gregory and S.~Ross, {\it Lifshitz flows in IIB and dual field theories}, JHEP  1411,(2014) 073 [ arXiv:1408.3271].
\bibitem{ho12} S. S. Gubser and A. Nellore, {\it Ground states of holographic superconductors}, Phys.Rev. D 80,(2009) 105007 [arXiv:0908.1972].
\bibitem{h12}  Y. C.  Ong and P.  Chen, {\it Stability of Horava-Lifshitz black holes in the context of AdS/CFT},  Phys. Rev. D 84, (2011) 104044 [arXiv:1106.3555].
\bibitem{h22}  M. Alishahiha and H.  Yavartanoo, {\it Conformally Lifshitz solutions from Horava-Lifshitz Gravity},  Class. Quant. Grav. 31, (2014) 095008 [arXiv:1212.4190].
\bibitem{oh12} S. Kachru, N. Kundu, A. Saha, R. Samanta and S. P. Trivedi, {\it Interpolating from Bianchi Attractors to Lifshitz and AdS Spacetimes}, JHEP 1403,(2014) 074 [ arXiv:1310.574].
\bibitem{d12} K.  Goldstein, N.  Iizuka, S. Kachru, S.  Prakash, S. P. Trivedi and A. {\it Westphal, Holography of Dyonic Dilaton Black Branes},  JHEP 1010, (2010) 027 [ arXiv:1007.2490].
\bibitem{d21} G. Bertoldi, B. A. Burrington and A. W. Peet, {\it Thermal behavior of charged dilatonic black branes in AdS and UV completions of Lifshitz-like geometries}, Phys. Rev. D 82, (2010) 106013 [ arXiv:1007.1464 ].
\bibitem{dh12} M. Kord Zangeneh, A. Sheykhi and  M. H. Dehghani, {\it Thermodynamics of topological nonlinear charged Lifshitz black holes}, Phys. Rev. D 92, (2015) 024050 [ arXiv:1506.01784].
\bibitem{hd12}  J. Tarrio and S. Vandoren, {\it Black holes and black branes in Lifshitz spacetimes}, JHEP 1109, (2011) 017 [arXiv:1105.6335].











\end{thebibliography}
\end{document}